%% file: stam_comp.tex
 \documentclass[12pt,onecolumn,twoside,letter]{IEEEtran}

\usepackage{url}
\usepackage{amsmath}
\usepackage{amsfonts}
\usepackage{theorem}
\usepackage[dvips]{epsfig}      
\usepackage{graphicx}
\usepackage{enumerate}
\usepackage{color}

\usepackage{verbatim}
\usepackage{amssymb}
\usepackage{amsbsy}
\usepackage{setspace}
\usepackage{url}

\newcommand{\bi}{\begin{itemize}}
\newcommand{\ei}{\end{itemize}}

\theoremstyle{plain}
\newtheorem{Theorem}{Theorem}
\newtheorem{Lemma}{Lemma}
\newtheorem{Proposition}{Proposition}
\newtheorem{Corollary}{Corollary}

 \newcommand{\Int}{\operatorname{Int}}

{\theorembodyfont{\rmfamily} }
{\theorembodyfont{\rmfamily} }
{\theorembodyfont{\rmfamily} \newtheorem{Example}{Example}}
{\theorembodyfont{\rmfamily} }
{\theorembodyfont{\rmfamily} }

\newcommand {\R}{\mathbb R}

\newcommand{\be}{\begin{equation}}
\newcommand{\ee}{\end{equation}}
\newcommand{\sgn}{\operatorname{{\mathrm sign}}}

\newcommand{\U}{\mathcal U}

\newcommand{\sname}{} \newcommand{\slabel}[1]{\debug{\fbox{\tiny \sname #1}}\label{\sname #1}}
\newcommand{\debug}[1]{}              
\newcommand{\FB}{\begin{figure}[t]\centering} \newcommand{\FE}[2]{\caption{#2 \debug{\fbox{\sname #1}}} \slabel{#1} \end{figure}} \newcommand{\tB}{\begin{table}[hbtp]\centering}
\newcommand{\tE}[2]{\caption{#2 \debug{\fbox{\sname #1}}}\slabel{#1} \end{table}} 

\newcommand{\st}{\, | \,}

\begin{document}
 \title{A Model for Competition for Ribosomes in the Cell \thanks{The research of MM and TT is partially supported by a research grant from  the Israeli Ministry of Science, Technology, and Space. The work of EDS is supported in part by grants AFOSR FA9550-14-1-0060 and ONR 5710003367.}}

\author{Alon Raveh, Michael Margaliot, Eduardo D. Sontag* and Tamir Tuller*\thanks{*Corresponding authors: EDS and TT.} \IEEEcompsocitemizethanks{\IEEEcompsocthanksitem
A. Raveh is with the School of Electrical Engineering, Tel-Aviv
University, Tel-Aviv 69978, Israel.
E-mail: ravehalon@gmail.com \protect\\
M. Margaliot is with the School of Electrical Engineering and the Sagol School of Neuroscience, Tel-Aviv
University, Tel-Aviv 69978, Israel.
E-mail: michaelm@eng.tau.ac.il \protect\\
E. D. Sontag  is with the
   Dept. of Mathematics and the   Center  for Quantitative Biology, Rutgers University, Piscataway, NJ 08854, USA.
	E-mail: eduardo.sontag@gmail.com \protect\\
T. Tuller is with the Dept. of Biomedical Engineering and the Sagol School of Neuroscience, Tel-Aviv
University, Tel-Aviv 69978, Israel.
E-mail: tamirtul@post.tau.ac.il  \protect\\

}}

\maketitle

\begin{abstract}

A single mammalian cell
includes an order of $10^4 - 10^5$  mRNA molecules and as many as $10^5 - 10^6$  ribosomes.
Large-scale simultaneous mRNA  translation and
the  resulting competition for the  available  ribosomes has important implications to the cell's functioning and evolution.
Developing a better understanding of the intricate correlations
 between  these simultaneous processes, rather than focusing
on the translation of a single  isolated transcript, should help in  gaining a better understanding of mRNA translation regulation and the way  elongation rates affect organismal fitness.
 A model of simultaneous translation is specifically important when dealing with highly expressed genes, as these consume more resources.
In addition, such a model  can lead
to more accurate predictions  that are needed in the interconnection of translational modules in  synthetic biology.

We develop and  analyze
 a general model  for large-scale  simultaneous mRNA translation and competition for ribosomes. This is based
on combining several  ribosome flow models~(RFMs) interconnected via
a pool of free ribosomes. We prove  that the compound system always converges to a steady-state and that it always entrains or phase locks to  periodically time-varying transition rates in any of the mRNA molecules.
We use this model to explore the interactions between the various mRNA molecules  and ribosomes at steady-state.
We show that increasing the length of an mRNA molecule decreases the production rate of all the mRNAs.
Increasing any of the codon translation rates in a specific mRNA molecule yields a local effect:
an
 increase in  the translation rate of this  mRNA,  and     also   a global effect:
the translation rates in the other mRNA molecules all increase or all decrease.
These  results suggest   that the effect of codon decoding rates of endogenous and heterologous mRNAs on protein production is more complicated than previously  thought.

\end{abstract}


\section{Introduction}

Various
processes in  the cell utilize the same  finite pool of available resources. This means that
the processes actually compete for these resources, leading to an indirect coupling between
the processes.  This is particularly relevant when many identical intracellular processes, all using
 the same resources,
 take place in parallel.

Biological evidence suggests that
  the  competition  for RNA polymerase (RNAP)
	and   ribosomes, and various transcription and translation factors, 
  is a  key  factor  in    the cellular economy of gene expression.
  The limited availability of these resources
  is one of the reasons why the levels of genes, mRNA, and proteins produced in the cell
  do not necessarily correlate~\cite{Tuller2010c,Kudla2009,Sharp2010,Tuller2015,Richter1981,peder1993,Jens2015}.

\begin{figure*}[t]
  \begin{center}
  \includegraphics[trim=0cm 12cm 0cm 0cm,scale=.55]{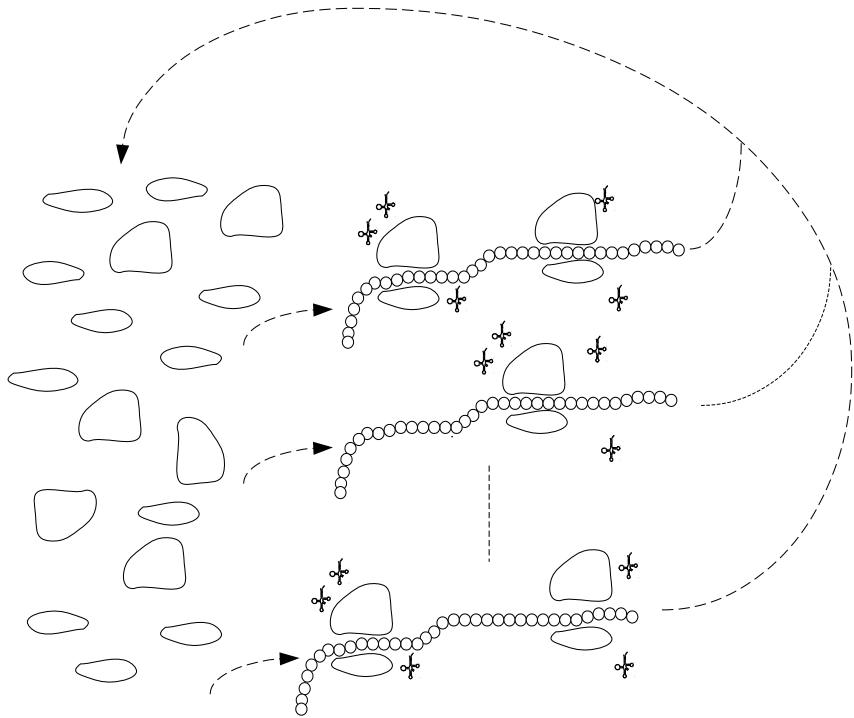}
  \caption{ Illustration of simultaneous translation of mRNA chains (right)  interconnected via
	a pool of free ribosomes (left). }\label{fig:simu}
  \end{center}
\end{figure*}

It was estimated that in  a yeast cell there are approximately  $60,000$
 mRNA  molecules. These can be translated in parallel~\cite{Zenklusen2008, Warner1999},
with possibly many ribosomes scanning the same transcript concurrently.
The number of ribosomes is limited  (in a yeast cell it is approximately~$ 240,000$) and
this leads to a competition
for ribosomes.
For example, if more ribosomes
bind to   a certain mRNA molecule  then the pool of free ribosomes in the cell is depleted,
and  this may lead to lower  initiation rates in the  other mRNAs  (see Fig.~\ref{fig:simu}).

There is a  growing interest in computational or mathematical
models that take into account
the competition for available resources in translation and/or transcription
(see, for example,~\cite{7040238,Gyorgy2015,Mather2013,Brewster2014,Dana2011,romano_2012_comp}).
One such model, that
   explicitly considers the movement of the ribosomes [RNAP] along  the mRNA [DNA],    is
  based on a set of
   \emph{asymmetric simple exclusion processes}~(ASEPs) interconnected to a pool of free
	ribosomes. ASEP  is an important model from non-equilibrium
	statistical physics
describing  particles that hop randomly from one site to the next
along an ordered  lattice  of  sites, but only  if the next site is empty.
This form of ``rough exclusion'' models the fact that the particles cannot overtake one another.
  ASEP has been used to model and analyze numerous multiagent systems with local interactions
  including the flow of  ribosomes along the mRNA molecule  \cite{MacDonald1968},\cite{Shaw2003}.
	In this context, the lattice represents the mRNA molecule,
	and the particles are the ribosomes.
	For more on  mathematical and computational models of translation, see the survey paper~\cite{haar_survey}.

Ref.~\cite{parking_garage_2002} considered  a closed  system composed
of a single  ASEP connected to a  pool (or reservoir) of ``free'' particles. The total number of particles is conserved. This is sometimes referred to as the \emph{parking garage problem},
with   the lattice    modeling  a road, the particles are cars,
and the pool corresponds to  a parking garage.
Ref.~\cite{Tasep_feed2009} studied a similar system using  domain wall theory.
Ref.~\cite{TASEPcomp12} (see also~\cite{mixed_tasep}) considered  a network composed of two ASEPs
connected to a finite pool of particles.
The analysis in these papers focuses on the  phase diagram of the compound system
with respect to certain parameters, and
on how the  phase of one  ASEP affects the phase of the other ASEPs.
These studies rely on the phase diagram of a single
ASEP that is well-understood only in the case where all the transition rates inside the chain (the elongation rates)
are equal. Thus, the network is typically  composed  of \emph{homogeneous}  ASEPs.
Another model~\cite{romano_2012_comp} combines  non-homogenous ASEPs in order to study
competition between  multiple species of
mRNA molecules  for a   pool of tRNA molecules.
This study was based on the \emph{Saccharomyces cerevisiae} genome.
However, in this case (and similar models, such as~\cite{Dana2011})  analysis seems intractable and one must resort
to  simulations only.

Our approach is based on
the \emph{ribosome flow model}~(RFM)~\cite{reuveni}. This  is a
 deterministic, continuous-time, synchronous model for translation that can be derived via
the mean-field approximation of ASEP~\cite{solvers_guide}.
The~RFM  includes $n$ state-variables describing the
ribosomal density in~$n$ consecutive sites along the mRNA molecule,
and~$n+1$ positive parameters: the initiation rate~$\lambda_0$,
 and the elongation rate~$\lambda_i$ from site~$i$ to site~$i+1$, $i=1,\dots,n$.
	 The~RFM  has
	a unique equilibrium point~$e=e(\lambda_0,\dots,\lambda_n)$, and  any trajectory emanating
  from a feasible initial condition converges to~$e$~\cite{RFM_stability} (see also~\cite{RFM_entrain}).
	This means that the system always converges to a steady-state ribosomal  density
	that depends on the rates, but not on the initial condition.
	In particular, the production rate converges to a steady-state value~$R$.
	The mapping from the rates to~$R$   is
    a concave function,
		so maximizing~$R$ subject to a  suitable constraint on the rates  is
	a convex optimization problem~\cite{RFM_concave,HRFM_concave}.
  Sensitivity analysis of the RFM with respect to the rates has been studied in~\cite{RFM_sense}.
 These results are  important in the context of  optimizing the protein production rate in synthetic biology.
Ref.~\cite{RFM_entrain} has shown that when the rates~$\lambda_i$ are time-periodic functions, with a common minimal period~$T$,
then every state-variable converges to a periodic  solution with period~$T$. In other words,
the ribosomal densities \emph{entrain}
to periodic excitations in the rates (due e.g. to periodically-varying abundances of tRNA molecules).

 In ASEP with \emph{periodic boundary conditions}
 a particle that hops from the last site returns
  to the first one. The mean field approximation of this model
  is  called the \emph{ribosome flow model on a ring}~(RFMR).
	The periodic boundary conditions mean that the total number of ribosomes is conserved.
	Ref.~\cite{RFMR}  analyzed the RFMR  using the theory of monotone
  dynamical systems that admit a first integral.

 Both  the  RFM and the RFMR  model mRNA translation on a  single mRNA molecule.
In this paper, we
introduce a new model, called the
 \emph{RFM network with a pool}~(RFMNP), that includes
  a network of~RFMs,  interconnected through
a dynamical   pool of free ribosomes, to model
and analyze simultaneous translation and competition for ribosomes in the cell.
To the best of our knowledge, this is the first study of a network of RFMs.
The total number of ribosomes in the RFMNP is conserved, leading to a
 first integral of the dynamics.
Applying  the theory of monotone dynamical systems that  admit a first integral we
prove several mathematical properties of the RFMNP: it    admits a continuum  of
equilibrium points, every trajectory converges to an equilibrium point,
and any two solutions emanating from initial conditions corresponding to an  equal number of ribosomes converge to the
same equilibrium point. These results hold for any set of rates  and in particular when the RFMs in the network
are not necessarily homogeneous.
These stability results are important  because they provide a rigorous framework
for studying  questions such as how does a change in one RFM   affects the
  behavior of all the other RFMs in the network? Indeed, since a steady-state
exists,  this can be reduced to asking how does a change in one RFM in the network  affects the
\emph{steady-state} behavior of the network?

To analyze  competition for ribosomes, we consider the effect of increasing one of the rates
in one~RFM, say RFM~\#1. This means that the ribosomes traverse RFM~\#1 more quickly.
 We show that this always leads to an increase in the production rate of RFM~\#1.
All the other RFMs are always affected in the same manner, that is, either all the other production rates
increase  or they all decrease.  
Our analysis shows that this can be explained as follows.
Increasing the rate~$\lambda_i$ in RFM~$\#1$
tends to increase   the steady-state
density in sites~$i+1,i+2,\dots$, and decrease the density in site~$i$  of this RFM.
The \emph{total} density (i.e., the sum of all the densities on the different sites along  RFM~$\#1$)
 can either decrease or increase. In the first case, more ribosomes are freed
to  the pool, and this increases the initiation
 rates
in all the other RFMs leading to higher production rates.
The second case leads to the opposite result.
The exact outcome of increasing one of the rates thus
depends  on  the many parameter values defining
 the pool and the set of  RFMs in the network.

 Our model takes into account  the  dynamics of the translation elongation stage, yet  is still amenable to analysis.
	This allows to develop a rigorous understanding of the effect of competition for ribosomes.
 Previous studies on this topic were either based on simulations (see, for example, \cite{Dana2011,romano_2012_comp})
or did not include a  dynamical model of translation elongation (see, e.g., \cite{Gyorgy2015,Brewster2014}).
For example, in an interesting paper, combining
mathematical modeling and biological experiments, Gyorgy et al.~\cite{Gyorgy2015}
study the expression levels of two adjacent reporter genes on a plasmid in {\em E. coli}  based on measurements of fluorescence levels, that is,  protein levels. These  are of course  the
 result of all the gene expression steps (transcription, translation, mRNA degradation, protein degradation)  making it difficult
  to  separately study  the effect of competition for ribosomes or to study specifically the translation elongation step.
  Their analysis  yields that the attainable output $p_1, p_2$ of the two proteins satisfies the formula
  \be\label{eq:dom}
\alpha p_1+\beta p_2 =Y,
  \ee
  where~$Y$ is related to the total number of ribosomes (but also other translation factors 
	and possibly additional gene expression  factors), and~$\alpha, \beta$ are constants that depend on
  parameters such as the plasmid copy number, dissociation constants of the ribosomes binding to the Ribosomal Binding Site (RBS), etc.
  This equation implies  that increasing the production of one protein always leads to a decrease in the production of the other protein
  (although more subtle correlations may take place via the effects on the constants~$\alpha$ and~$\beta$).
  A similar conclusion has been derived for other models as well~\cite[Ch.~7]{delmurrayboo}.

In our model,  improving the translation rate of a codon  in
one mRNA   may either   increase or decrease  the translation rates of \emph{all} other mRNAs in the cell.
Indeed, the effect on the other genes depends on the change in the \emph{total density} of ribosomes on the modified mRNA molecule, highlighting
the importance of modeling the dynamics of the translation elongation step. 
We show however that when increasing the initiation rate in an RFM in the network, 
the total density in this RFM always increases and, consequently,
the production rate in all the other RFMs decreases. This special case agrees
with the results in~\cite{Gyorgy2015}.

Another recent study~\cite{counter_vecc} showed that the hidden layer
 of interactions among genes  arising from competition for shared resources
can dramatically change network
behavior. For example, a cascade of activators can behave like an
effective repressor, and a repression
cascade can become bistable. This agrees with several
 previous studies in the field (see, for example, \cite{Jens2015,Ala2013}).

The remainder of this paper is organized as follows.
The next section describes the new  model.
We demonstrate using several examples how it can be used to
study  translation at the cell level.
Section~\ref{sec:main} describes our main theoretical  results, and details  their biological implications.
To streamline the presentation, the proofs of the  results are placed in the Appendix.
The final section concludes and describes several directions
 for further research.

\section{The  model and some examples}\label{sec:model}
Since our model is based on a network of interconnected  RFMs, we begin with a brief review of the~RFM.
\subsection{Ribosome flow model}
The  RFM  models the traffic flow of ribosomes along the mRNA.
 The mRNA chain is divided into a set of~$n$ compartments or  {sites}, where each site may correspond
to  a codon or a group of codons.
The state-variable~$x_i(t)$, $i=1,\dots,n$,   describes  the  ribosome occupancy at site~$i$ at time~$t$, normalized
such that
 $x_i(t)=0$ [$x_i(t)=1$] implies that site~$i$ is completely empty [full] at time~$t$.
Roughly speaking, one may also view~$x_i(t)$ as the probability that site~$i$ is occupied at time~$t$.
The dynamical equations of the RFM are:
\begin{align}\label{eq:just_rfm}
                    \dot{x}_1&=\lambda_0 (1-x_1) -\lambda_1 x_1(1-x_2), \nonumber \\
                    \dot{x}_2&=\lambda_{1} x_{1} (1-x_{2}) -\lambda_{2} x_{2} (1-x_3) , \nonumber \\
                    \dot{x}_3&=\lambda_{2} x_{ 2} (1-x_{3}) -\lambda_{3} x_{3} (1-x_4) , \nonumber \\
                             &\vdots \nonumber \\
                    \dot{x}_{n-1}&=\lambda_{n-2} x_{n-2} (1-x_{n-1}) -\lambda_{n-1} x_{n-1} (1-x_n), \nonumber \\
                    \dot{x}_n&=\lambda_{n-1}x_{n-1} (1-x_n) -\lambda_n x_n   .
\end{align}
These equations describe the movement of ribosomes along the mRNA chain.
The \emph{transition rates}~$\lambda_0,\dots,\lambda_n$ are
all positive numbers (units=1/time).
To explain this model, consider the equation~$\dot x_2=\lambda_{1} x_{1} (1-x_{2}) -\lambda_{2} x_{2} (1-x_3)$. The term~$ \lambda_{1} x_{1} (1-x_{2})$ represents the flow
of particles from site~$1$ to site~$2$. This is proportional to the occupancy~$x_1$ at site~$1$
and also to~$1-x_2$, i.e. the flow decreases as site~$2$ becomes fuller. In particular, if~$x_2=1$, i.e. site~$2$
is completely full, the flow from site~$1$ to site~$2$ is zero. This is a
``soft''  version of the rough  exclusion principle in ASEP.
Note that the maximal possible flow rate from site~$1$ to site~$2$ is the transition rate~$\lambda_1$.
The term~$ \lambda_{2} x_{2} (1-x_3)$   represents the flow
of particles from site~$2$ to site~$3$.

\begin{figure}[t]
 \centering
 \scalebox{1}{\input{fig_rfm_l0.pstex_t}}
\caption{  Topology of the  RFM.  State variable~$x_i(t)\in[0,1]$ describes
 the normalized
ribosome occupancy level in site~$i$ at time $t$. The    initiation rate is~$\lambda_0$,  and
 $\lambda_i$  is the elongation rate  between sites~$i$ and~$i+1$. Production rate at time $t$ is $R(t):=\lambda_n x_n(t)$.}\label{fig:rfmm}
\end{figure}
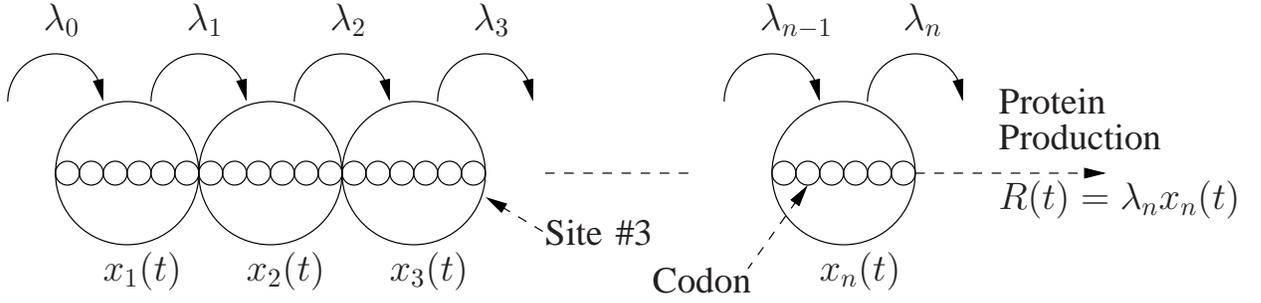

The dynamical  equations for the other state-variables are  similar. Note that~$\lambda_0$ controls the initiation
 rate into the chain, and that
\[
            R(t):=\lambda_n x_n (t),
\]
is the rate of flow of ribosomes out of the chain, that is,
   the translation   (or protein production) rate at time~$t$.
The RFM topology is depicted in Fig.~\ref{fig:rfmm}.

The RFM encapsulates simple exclusion and unidirectional movement  along the lattice
just as in
 ASEP.
  This is not surprising, as the RFM  can be derived via a mean-field approximation
of ASEP
(see, e.g., \cite[p.~R345]{solvers_guide} and~\cite[p.~1919]{PhysRevE.58.1911}).
However, the analysis of these two models is quite different as the RFM is
  a deterministic, continuous-time, synchronous model, whereas ASEP
	is a stochastic, discrete-type, asynchronous one.

In order to study a  network of interconnected  RFMs, it is useful to first extend the  RFM
  into  a single-input single-output (SISO)
control system:
\begin{align}\label{eq:rfmio}
                    \dot{x}_1&= \lambda_{0} (1-x_1)u -\lambda_1 x_1(1-x_2), \nonumber \\
                    \dot{x}_2&=\lambda_{1} x_{1} (1-x_{2}) -\lambda_{2} x_{2} (1-x_3) , \nonumber \\
                    \dot{x}_3&=\lambda_{2} x_{ 2} (1-x_{3}) -\lambda_{3} x_{3} (1-x_4) , \nonumber \\
                             &\vdots \nonumber \\
                    \dot{x}_{n-1}&=\lambda_{n-2} x_{n-2} (1-x_{n-1}) -\lambda_{n-1} x_{n-1} (1-x_n), \nonumber \\
                    \dot{x}_n&=\lambda_{n-1}x_{n-1} (1-x_n) -\lambda_n x_n  ,\nonumber\\
                    y&=\lambda_n x_n.
\end{align}
\textcolor{black}{
Here the  translation rate becomes the output~$y$ of the system,
and  the flow into site~$1$ is multiplied  by a time-varying
control~$u:\R_+\to \R_+$, representing the flow of ribosomes from the ``outside world'' into the strand which is related to the rate  ribosomes diffuse to the 5'end (in eukaryotes) or the RBS (in prokaryotes) of the mRNA.
Of course,  mathematically one can absorb~$\lambda_0$ into~$u$, but we do not do this because we think of~$\lambda_0$
as representing some  local/mRNA-specific features of the mRNA sequence (e.g. the strength of the Kozak sequences in eukaryotes or the  RBS  in prokaryote).
}

The set of admissible controls~$\U$ is the set of bounded and measurable functions  taking values in~$\R_+$ for  all time~$t\geq0$.
 Eq.~\eqref{eq:rfmio}, referred to as the RFM with input and output (RFMIO)~\cite{RFM_feedback},
  facilitates  the study of RFMs with feedback connections.
We note in passing that~\eqref{eq:rfmio} is a \emph{monotone control system} as defined in~\cite{mcs_angeli_2003}.
From now on we write~\eqref{eq:rfmio} as
\begin{align}\label{eq:network}
            \dot{x} &= f( x, u ),\nonumber \\
                    y &=\lambda_n x_n .
\end{align}

Let
\[
           C^n:=\{z \in \R^n: z_i \in [0,1] ,\; i=1,\dots,n\},
\]
denote the closed unit
cube in~$ \R^n$.
Since the state-variables represent  normalized occupancy  levels,
we always consider initial conditions   $x(0)\in C^n$.
 It is straightforward to verify that~$C^n$ is an invariant set of~\eqref{eq:network},
i.e. for any~$u\in\U$ and any~$x(0)\in C^n$ the trajectory satisfies~$x(t,u) \in C^n$ for all~$t\geq 0$.

 \subsection{RFM network with a pool}
\begin{figure*}[t]
  \begin{center}
  \input{fig_rfm_network_with_pool.pstex_t}
  \caption{Topology of the RFMNP.} \label{fig:rfm_network}
  \end{center}
\end{figure*}
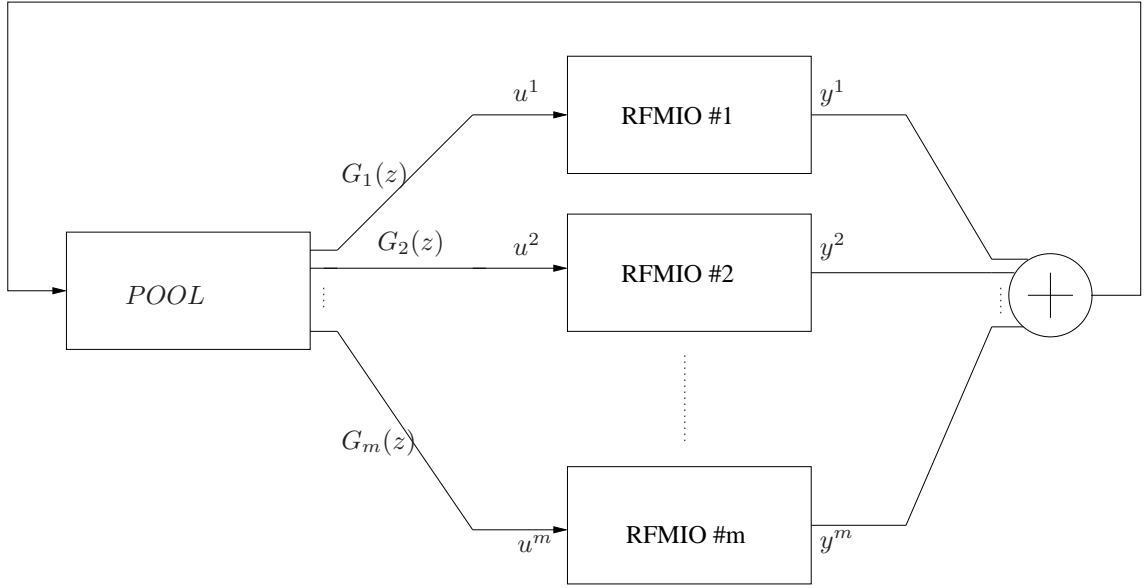

To model competition for ribosomes in the cell, we consider a set of~$m\geq 1 $ RFMIOs, representing~$m$ different
mRNA chains.
The $i$th RFMIO has length~$n_i$, input function~$u^i$, output function~$y^i$,
 and rates~$\lambda^i_0,\dots,\lambda^i_{n_i}$.
The RFMIOs are
interconnected through a  pool of free ribosomes (i.e., ribosomes that are not attached to any mRNA molecule).
The output of each  RFMIO  is fed into the pool, and
the pool feeds the initiation locations in the  mRNAs (see Fig.~\ref{fig:rfm_network}). Thus, the model includes~$m$ RFMIOs:
\begin{align}\label{eq:nolin}
                            \dot x^1&= f(x^1,u^1), \quad y^1=\lambda^1_{n_1} x_{n_1} ,\nonumber\\
                            \vdots \nonumber\\
                            \dot x^m&= f(x^m,u^m),  \quad y^m=\lambda^m_{n_m} x_{n_m},
\end{align}
and a dynamic  pool of ribosomes   described    by
\begin{align}\label{eq:pool}
           \dot z =\sum_{j=1}^m y^j - \sum_{j=1}^m \lambda^j_0 (1-x_1^j)G_j(z).
\end{align}
where~$z:\R_+\to \R$ describes  the pool occupancy.
Eq.~\eqref{eq:pool} means that the flow into the pool is the sum of all
output rates~$\sum_{j=1}^m y^j $
 of the RFMIOs minus the total flow of ribosomes that bind to an mRNA molecule~$\sum_{j=1}^m   \lambda^j_0 (1-x_1^j)G_j(z)$.
 Recall that  the term~$(1-x_1^j)$ represents  the exclusion, i.e. as the first site in RFMIO~$\#j$ becomes fuller,
 less ribosomes can bind to it.   Thus, the input to RFMIO~$\#j$ is
 \be\label{eq:inputj}
 u^j(t)= G_j(z
(t)),\quad j=1,\dots,m.
 \ee
We   assume that each $G_j(\cdot):\R_+ \to \R_+ $ satisfies: (1)~$G_j(0)=0$;
 $G_j$ is~continuously differentiable and~$G_j'(z)>0$ for all~$z\geq 0$ (so~$G_j $ is strictly increasing on~$\R_+$);
and (3)~there exists $s>0$ such that~$G_j(z)\leq s z$ for all~$z>0$
sufficiently small.
These properties imply in particular that if the pool is empty then no ribosomes can bind to the mRNA chains, and
that as the pool becomes fuller the initiation rates to the RFMIOs increase.
 
Typical examples for functions satisfying these properties
include  the linear function, say,~$G_j(z)=  z$,
and~$G_j(z)=a_j\tanh(b_jz)$, with~$a_j,b_j >0$. In the first case, the flow
of ribosomes into the first site of~RFM~$\#i$ is given by
$\lambda^i_0 z (1-x^i_1)$, and  the product here can be justified via  mass-action kinetics. The use of~$\tanh$
 may be suitable for
modeling a saturating function.
This is in fact  a standard function   in ASEP models with a pool~\cite{TASEPcomp12,const_tasep_08},
because it is zero when~$z$ is zero, uniformly bounded, and strictly increasing for~$z\geq0$.
Also, for~$z\geq0$ the function~$\tanh(z)$ takes values in~$[0,1)$ so
it can also be interpreted  as a probability function~\cite{TASEPopt11}.

 In the context of a shared pool, it is natural to consider
 the special case where~$G_j(z)=G(z)$ for all~$j=1,\dots,m$.
The differences between the initiation sites in the strands are then modeled by the different~$\lambda_0^j$'s.

Note that combining the  properties  of~$G_j$ with~\eqref{eq:pool}
implies that if~$z(0)\geq 0$ then~$z(t)\geq 0$ for all~$t\geq 0$. Thus, the pool occupancy is always non-negative.

Summarizing, the \emph{RFM network with a pool}~(RFMNP) is given  by equations~\eqref{eq:nolin},~\eqref{eq:pool},
and~\eqref{eq:inputj}. This is  a dynamical system with~$d:=1+\sum_{i=1}^m n_{i}$ state-variables.

\begin{Example}
Consider a network  with~$m=2$ RFMIOs, the first [second]  with dimension~$n_1=2$ [$n_2=3$]. Then the RFMNP is given by
\begin{align}\label{eq:exa223}
                    \dot{x}^1_1&= \lambda^1_0  (1-x_1^1)G_1(z)-\lambda_1^1 x_1^1(1-x_2^1), \nonumber \\
                    \dot{x}^1_2&=\lambda^1_{1} x^1_{1} (1-x^1_{2}) -\lambda_{2}^1 x_{2}^1  , \nonumber \\
                    \dot{x}_1^2&= \lambda^2_0 (1-x_1^2)G_2(z) -\lambda_1^2 x_1^2(1-x_2^2), \nonumber \\
                    \dot{x}_2^2&=\lambda_{1}^2 x_{1}^2 (1-x_{2}^2) -\lambda_{2}^2 x_{2}^2 (1-x_3^2) , \nonumber \\
                    \dot{x}_3^2&=\lambda_{2}^2 x_{ 2}^2 (1-x_{3}^2) -\lambda_{3}^2 x_{3}^2 , \nonumber \\
                     \dot z &= \lambda_{2}^1 x_{2}^1 + \lambda_{3}^2 x_{3}^2 - \lambda^1_0   (1-x_1^1)G_1(z)-  \lambda^2_0  (1-x_1^2)G_2(z).
\end{align}
Note that this system has~$d=6$ state-variables.~$\square$
\end{Example}

An important property of the RFMNP is, that being a closed system,
the total occupancy
\be\label{eq:nrivi}
         H(t):=  z(t) + \sum_{j=1}^m  \sum_{i=1}^{n_j}   x^j_i(t) ,
\ee
is conserved, that is,
\be\label{eq:Hfixed}
        H(t) \equiv  H(0), \quad \text{for all } t\geq 0.
\ee
In other words,~$H$ is a first integral of the dynamics.
In particular, this means  that~$z(t)\leq H(t) = H(0)$ for all~$t\geq 0$, i.e. the pool occupancy   is uniformly  bounded.

The RFMNP models mRNAs that  compete  for ribosomes because the total number of ribosomes is conserved.
As more ribosomes bind to  the RFMs, the pool depletes,~$G_j(z)$ decreases,
and the effective initiation rate to all the RFMs decreases.
 This allows  to systematically address
  important   biological  questions on large-scale simultaneous translation under competition for ribosomes.
 The following  examples demonstrate this.
 We prove in Section~\ref{sec:main}
  that all the state-variables in the RFMNP converge to a steady-state.
	Let~$e^i_j \in[0,1]$ denote the steady-state occupancy in site~$j$ in RFMIO~$\#i$, and let~$e_z \in[0,\infty)$
	denote the steady-state occupancy in the pool.
In the examples below we always consider these steady-state values (obtained
numerically by simulating the differential equations).

\begin{figure*}[t]
  \begin{center}
  \includegraphics[scale=.6]{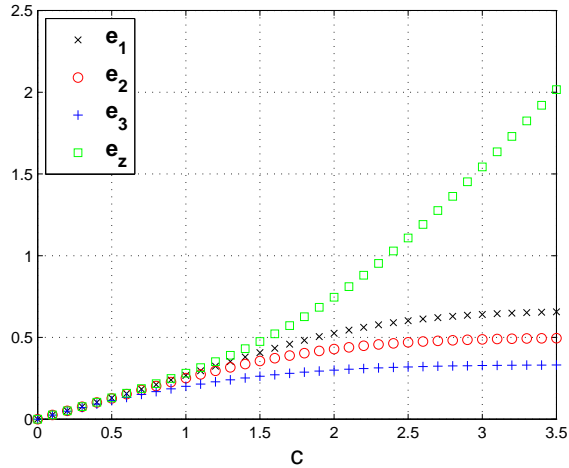}
  \caption{Steady-state values in the RFMNP in Example~\ref{exa:onestrand} as a function of the total occupancy~$H(0)=c$. }\label{fig:sinrm}
  \end{center}
\end{figure*}

\begin{Example}\label{exa:onestrand}
Although we are mainly interested in modeling large-scale simultaneous translation, it is natural to first consider
a model with a single mRNA molecule  connected to a pool of ribosomes.
From a biological perspective, this models   the case where there is one
 gene that is highly expressed with respect  to all other genes
(e.g. an extremely highly expressed heterologous gene).

Consider an  RFMNP  that includes a single RFMIO (i.e.~$m=1$),
with dimension~$n_1=3$,   rates~$\lambda^1_i=1$, $i= 0,1,2,3$,
and a pool with output function~$G(z)=\tanh(z)$.
We simulated this system for the initial condition~$x^1_i(0)=0$, $z(0)=c$ for various values of~$c$.
Note that~$H(t)\equiv H(0)=c$.
Fig.~\ref{fig:sinrm}
depicts the steady-state values~$e_1,e_2,e_3$ of the state-variables in the RFMIO,
 and the steady-state pool occupancy~$e_z$.
It may be seen that for small values of~$c$ the steady-state
ribosomal densities and thus the production rates are very low.
This is simply because there are not enough ribosomes in the network.
The ribosomal densities increase with~$c$. For large values of~$c$, the output function of the pool saturates, as $\tanh(z)\to 1$,
and so does
the initiation rate in the RFMIO. Thus, the  densities in the RFMIO saturate to the values corresponding to
the initiation rate~$\lambda_0=1$,  and then
all the remaining ribosomes accumulate in the pool. Using a different pool output function, for example~$G(z)=z$, leads to the same
qualitative behavior, but with higher  saturation values for the ribosomal  densities in the RFMIO.
(Note that the ribosomal densities in an RFM are finite even when~$\lambda_0\to\infty$~\cite{HRFM_steady_state}.)~$\square$
\end{Example}

This  simple
example already demonstrates the coupling between the ribosomal pool, initiation rate, and elongation rates.
  When the ribosomal pool is small the initiation rate is   low. Thus,
  the ribosomal densities on the mRNA are low and there are no interactions between ribosomes (i.e., no ``traffic jams'') along the~mRNA.
  The initiation rate becomes the rate limiting step of translation.
   On the other-hand, when there are many ribosomes in the pool
 the initiation rate increases, the elongation rates become rate limiting and ``traffic jams''
 along the mRNA  evolve. At some point, a further increase  in the number of ribosomes in the pool
 will have a negligible  effect on the production rate.

It is known  that there can be very large changes  in the number of ribosomes in the cell during e.g.
 exponential growth. For example. Ref.~\cite{Bremer1996} reports
changes in the range $6,800$ to $72,000$. The example above demonstrates
how these large changes in the number of ribosomes are expected to affect the translational regimes; specifically, it may cause a  switch between the different
regimes mentioned above.

\begin{figure*}[t]
  \begin{center}
  \includegraphics[scale=.6]{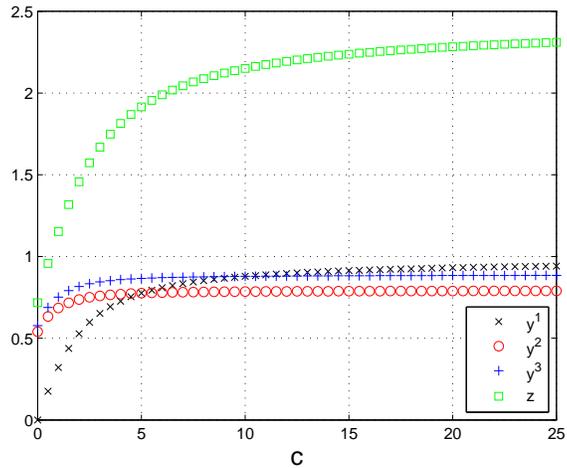}
  \caption{Steady-state outputs~$y^i$
   of the  three RFMIOs and pool occupancy~$z$  as a function of the homogeneous transition rate~$c$ in  RFMIO~$\#1$. }\label{fig:allc}
  \end{center}
\end{figure*}

\begin{figure*}[t]
  \begin{center}
  \includegraphics[scale=.6]{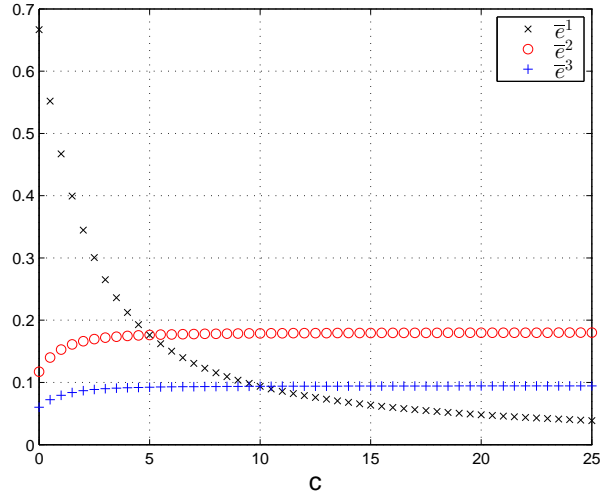}
  \caption{Average steady-state occupancy
   in  the  three RFMIOs as a function of the homogeneous transition rate~$c$ in  RFMIO~$\#1$. }\label{fig:allm}
  \end{center}
\end{figure*}

The next example describes an RFMNP with several  mRNA chains.
Let~$1_n \in \R^n$ denote the vector of~$n$ ones.
\begin{Example}\label{exp:current}
Consider an  RFMNP  with~$m=3$ RFMIOs of dimensions~$n_1=n_2=n_3=3$,  and    rates
\begin{align*}
		     \lambda^1_i = c ,  \quad
		     \lambda^2_i = 5,  \quad
		     \lambda^3_i = 10,\quad i=0,\dots,3.
\end{align*}
 In other words, every RFMIO has homogeneous rates. Suppose  also that~$G_i(z)=\tanh(z)$, for~$i=1,2,3$.
We simulated this RFMNP for different values of~$c$ with the initial condition~$z(0)=0$,
$x^1(0)=(1/2)1_3$,
$x^2(0)=(1/3)1_3$ and~$x^3(0)=(1/4)1_3$.
Thus,~$H(0)=3.25$   in all the simulations.
For each value of~$c$, every state-variable in the RFMNP converges to a steady-state.
Fig.~\ref{fig:allc} depicts the steady-state value~$e_z$ and the steady-state
output~$y^i$ in each RFMIO.
It may be seen that increasing~$c$, i.e. increasing all the elongation rates in RFMIO~$\#1$
leads to an increase in the steady-state translation rates in \emph{all} the~RFMIOs in the network.
Also, it leads to an increase in the steady-state occupancy of the pool.
It may seem that this contradicts~\eqref{eq:Hfixed}
but this is not so. Increasing~$c$ indeed increases all the steady-state translation rates,
but it decreases the steady-state occupancies inside each RFMIO
so that  the total~$H(t)=H(0)=3.25$ is conserved.

Define the \emph{average steady-state occupancy}~(ASSO) in  RFMIO $\#j$    by
$\bar e^j:=  \frac{1}{n_j}\sum_{i=1}^{n_j}e^j_i$.
Fig.~\ref{fig:allm} depicts the ASSO in each RFMIO
 as a function of~$c$.
It may be seen as~$c$ increases the ASSO in RFMIO~$\#1$ decreases quickly,
yet  the ASSOs in the other two RFMIOs slowly increases.  Since the ribosomes spend less time on RFMIO~$\#1$ (due to increased~$c$) they are now available for translating the other RFMIOs, leading  to the increased ASSO in the  other mRNAs.~$\square$
\end{Example}

From a biological point of view this example corresponds
to a situation where accelerating \emph{one} of the~mRNA chains
increases the protein production rates in \emph{all} the mRNAs and also increases the number of free ribosomes.
Surprisingly, perhaps, it also suggests  that a relatively
  larger number of free ribosomes in the cell corresponds to higher protein production rates. This agrees
	with   evolutionary, biological, and synthetic biology studies that have suggested that (specifically) highly expressed genes (that are transcribed into many mRNA molecules) undergo selection to include codons with improved elongation rates~\cite{Kudla2009,Tuller2010,Tuller2010c}.
	Specifically,  two mechanisms by which improved codons affect translation efficiency and the organismal fitness are~\cite{Tuller2010}:
	(1)~\emph{global mechanism}: selection for improved codons contributes toward improved ribosomal recycling and global allocation; the increased number of free ribosomes improves
	the effective translation initiation rate of all genes, and thus  improves global translation efficiency;
	and (2)~\emph{local mechanism}:
	the improved translation elongation rate of an mRNA contributes directly to its protein production rate.

  The example  above demonstrates  both mechanisms, as  improvement of the translation elongation rates of one RFM increases the translation rate of this mRNA (local translation efficiency), and also of the other RFMs (global translation efficiency).
  In addition, as can be seen,  the decrease in ASSO in RFMIO~$\#1$ is significantly higher than the increase in ASSO in the other RFMIO. Thus, the simulation also demonstrates that increasing  the translation rate $c$ may
	contribute to decreasing ribosomal collision (and possibly ribosomal abortion).

We prove in Section~\ref{sec:main}  that when one of the rates in one of the RFMIOs increases
two outcomes are possible: either all the production rates in the other RFMs increase (as in this example)
or  they all decrease.
 As discussed below,  we believe that this second case is less likely to occur in endogenous genes, but may occur in heterologous gene expression.

\begin{figure*}[t]
  \begin{center}
  \includegraphics[scale=.6]{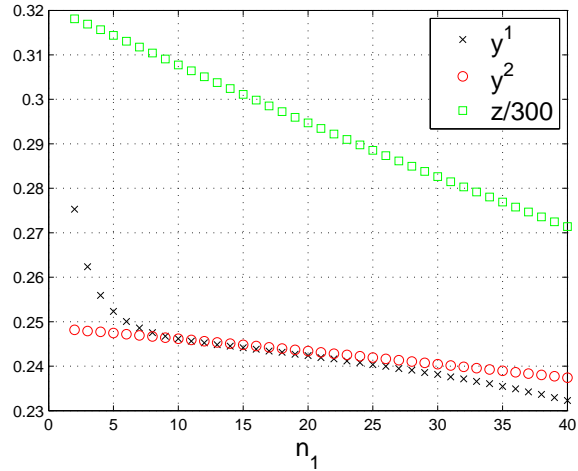}
  \caption{Steady-state outputs~$y^i$
   of the  two RFMIOs and steady-state  pool occupancy~$z/300$  as a function of the length~$n_1$ of  RFMIO~$\#1$.
   The normalization of~$z$ by $300$ is used to obtain similar magnitudes for all the values only.}\label{fig:longrfm}
  \end{center}
\end{figure*}

 The next example describes the effect of changing the length of one RFMIO in the network.
\begin{Example}\label{exa:long}
Consider an  RFMNP  with~$m=2$ RFMIOs of dimensions~$n_1$ and~$n_2=10$,    rates
\begin{align*}
		     \lambda^1_i &= 1 ,  \quad i=0,\dots,n_1,\\
		     \lambda^2_j &= 1,  \quad j=0,\dots, 10,
\end{align*}
and~$G_i(z)=\tanh(z/200)$,~$i=1,2 $. In other words, both RFMIOs have the same homogeneous   rates.
We simulated this RFMNP for different values of~$n_1$ with the initial condition~$z(0)=100$,
$x^1(0)=0_{n_1}$,
$x^2(0)=0_{10}$.
Thus~$H(0)= 100$ in all the simulations.
For each value of~$n_1$, every state-variable in the RFMNP converges to a steady-state.
Fig.~\ref{fig:longrfm} depicts the steady-state values of~$z$, and the steady-state
output~$y^i$ in each RFMIO.
It may be seen that increasing~$n_1$, i.e. increasing the length of RFMIO~$\#1$
leads to a decrease in the steady-state production rates and in the
steady-state pool occupancy. This is reasonable, as  increasing~$n_1$ means that
ribosomes that bind to the first chain remain on it for a longer period of time.
This decreases the production rate~$y^1$ and, by the competition for ribosomes,
also decreases the pool occupancy and thus decreases~$y^2$.~$\square$
  \end{Example}


From a biological point of view this suggests
 that decreasing the length of mRNA molecules contributes locally and globally to improving translation efficiency. A shorter coding sequence   improves the translation rate  of the mRNA and, by competition, may also
  improve the translation rates in all other mRNAs. Thus, we should expect to see selection for shorter coding sequences, specifically in highly expressed genes and in organisms with large population size. Indeed, previous studies have reported that in some organisms the coding regions of highly expressed genes tend to be shorter \cite{Dana2011}; other studies have
	shown that other (non-coding) parts of highly expressed genes tend to be shorter \cite{Castillo2002,Eisenberg2003,Li2007,Tuller2009}.
  Decreasing the length of different parts of the gene should contribute to organismal fitness via improving the energetic cost of various gene expression steps. For example, shorter genes should improve the metabolic cost of synthesizing mRNA and proteins; it can also reduce  the energy spent for splicing and processing of RNA and proteins. However, there are of course
   various functional and regulatory constraints that also contribute to shaping the gene length (see, for example \cite{Carmel2009}).
Our results and these previous studies suggest that in some cases genes are expected to undergo selection also for short {\em coding}  regions, as this reduces the required number of translating ribosomes.

  The next section describes
  theoretical results on the~RFMNP. All the proofs are placed in the Appendix.
 \section{Mathematical properties of the RFMNP}\label{sec:main}
Let
  \[
    \Omega:=[0,1]^{n_1} \times \dots \times [0,1]^{ n_m} \times [0,\infty)
  \]
  denote the state-space of the RFMNP (recall that every~$x_i^j$ takes values in $[0,1]$
	and~$z\in [0,\infty)$).  For an initial condition~$a \in \Omega$, let~$\begin{bmatrix} x(t,a)&z(t,a)\end{bmatrix}'$ denote the solution of the~RFMNP at time~$t$.
It is straightforward to show that  the solution remains in~$\Omega$ for all~$t\geq 0$.
Our first result shows that a slightly stronger property holds.
\subsection{Persistence}
\textcolor{black}{
\begin{Proposition}\label{prop:inv}
    For any~$\tau>0$ there exists~$\varepsilon=\varepsilon(\tau)>0$,
		with~$\varepsilon(\tau) \to 0$ when~$\tau\to 0$, such that for all~$t\geq \tau$, all~$j=1,\dots, m $, all~$i=1,\dots,n_j$,
		and all~$a\in (\Omega\setminus\{0\})$,
    \[
                \varepsilon \leq x^j_i(t,a)  \leq 1- \varepsilon  ,
    \]
		and
			\[
                \quad \varepsilon \leq z(t,a).
    \]
\end{Proposition}}
In other words, after time~$\tau$ the solution is~$\varepsilon$-separated from the boundary of~$\Omega$. This result is useful
 because
  on the boundary of~$\Omega$, denoted~$\partial \Omega$, the RFMNP looses some  desirable properties. For example, its Jacobian matrix
may become   reducible on~$\partial \Omega$.  Prop.~\ref{prop:inv} allows us to overcome this technical difficulty, as it implies that
any trajectory is separated  from the  boundary after an arbitrarily  short time.

\subsection{Strong Monotonicity}
Recall that a cone~$K\subseteq \R^n$  defines a partial order  in~$\R^n$ as follows.
 For two vectors~$a,b \in \R^n$, we write~$a\leq b$ if~$(b-a) \in K$;
  $a<b$ if~$a\leq b$ and~$a \not =b$; and~$a \ll b$ if~$(b -a )\in \Int(K)$.
  A   dynamical system~$\dot{x}=f(x)$ is called
  \emph{monotone} if~$a \leq b$ implies that~$x(t,a)\leq x(t,b)$ for all~$t \geq 0$.
  In other words, monotonicity means that the  flow preserves the partial ordering~\cite{hlsmith}.
It is called \emph{strongly monotone} if~$a < b$ implies that~$x(t,a)\ll x(t,b)$ for all~$t > 0$.

From here on we consider
  the  particular case
  where the cone is~$K=\R^n_+$.
  Then~$a\leq b$ if~$a_i\leq b_i$ for all~$i$,
   and~$a \ll b$ if~$a_i <b_i$ for all~$i$.
   A system that is monotone with respect to this partial order is called \emph{cooperative}.

   The next result analyzes the cooperativity of the~RFMNP.
 Let~$d:={1+\sum\limits_{i=1}^m n_i}$ denote the dimension of the~RFMNP.
\begin{Proposition}\label{prop:mono}
For any~$a,b \in \Omega$ with~$a \leq b$,
                \be\label{eq:abab}
                            x(t,a) \leq x(t,b)  \text{ and }  z(t,a) \leq z(t,b), \quad \text{for all } t \geq 0.
                \ee
Furthermore, if~$a< b$ then
                \be\label{eq:strongabab}
                            x(t,a) \ll x(t,b)  \text{ and }  z(t,a) < z(t,b) , \quad \text{for all } t  > 0.
                \ee
\end{Proposition}

This means the following.
Consider the RFMNP initiated with two initial conditions  such that
 the ribosomal densities in every site and the pool
corresponding to the first initial condition
are smaller or equal to the densities in the second initial condition. Then this correspondence between the
densities remains true for all time~$t\geq0$.

\subsection{Stability}
For~$s \geq 0$,  let
\[
L_s:=\{ y\in \Omega:   1_d' y= s     \}.
\]
In other words,~$L_s$ is a \emph{level set} of the first integral~$H$.
\begin{Theorem} \label{thm:main}
Every level set~$L_s$, $s\geq 0$, contains a unique equilibrium point~$e_{L_s}$ of the RFMNP, and
for  any initial condition~$a \in L_s$, the solution of the RFMNP converges to~$e_{L_s}$.
Furthermore, for any~$0\leq s  < p  $,
\be\label{eq:linord}
e_{L_s} \ll e_{L_p}.
\ee
\end{Theorem}

In particular, this means that every trajectory converges to an
  an equilibrium point, representing steady-state ribosomal densities in the RFMIOs and the pool.
	Note that Proposition~\ref{prop:inv} implies that for any~$s>0$,~$e_{L_s} \in \Int(\Omega)$.
	Eq.~\eqref{eq:linord} means that
  the continuum of   equilibrium points, namely,
 $\{e_{L_s}:\; s \in [0,\infty)\}$, are linearly ordered.

\begin{figure*}[t]
  \begin{center}
  \includegraphics[scale=.6]{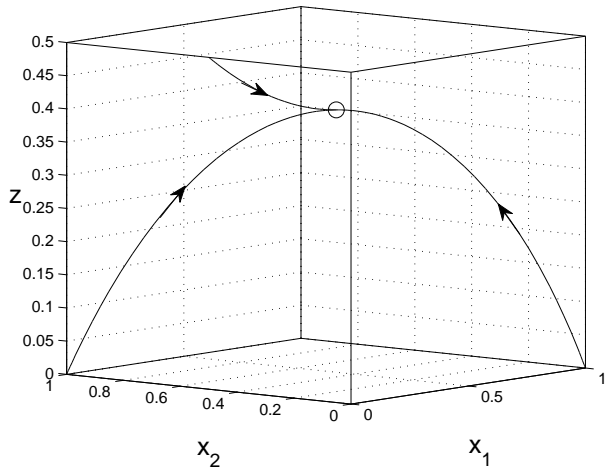}
  \caption{Trajectories of the system in Example~\ref{exa:3simp} for three initial conditions  in~$L_1$.
  The equilibrium point~\eqref{eq:sole3} is marked by a circle. }\label{fig:e1e2e3}
  \end{center}
\end{figure*}

\begin{Example}  \label{exa:3simp}
Consider an RFMNP with~$m=2$ RFMIOs with dimensions~$n_1=n_2=1$, and~$G_i(z)=z$, $i=1,2$, i.e.
\begin{align}\label{eq:simp_3}
                     \dot{x}^1_1&= \lambda_0^1 (1-x_1^1)z-\lambda_1^1 x_1^1, \nonumber \\
                    \dot{x}_1^2&=  \lambda_0^2 (1-x_1^2)z -\lambda_1^2 x_1^2, \nonumber \\
                     \dot z &= \lambda_{1}^1 x_{1}^1 + \lambda_{1}^2 x_{1}^2 - \lambda_0^1  (1-x_1^1)z-   \lambda_0^2  (1-x_1^2)z.
\end{align}
Note that even in this simple case the RFMNP is a nonlinear system.
Assume that~$\lambda_0^1 =\lambda_0^2 =1$, and that~$\lambda^1_1=\lambda^2_1$, and denote this value simply by~$\lambda$.
Pick an initial condition in~$\Omega$, and
let~$s:=x_1^1(0)+x_1^2(0)+z(0)$, so that the trajectory belong to~$L_s$ for all~$t\geq 0$.
Any equilibrium point~$e=\begin{bmatrix}e_1&e_2&e_z
\end{bmatrix} ' \in L_s$ satisfies
\begin{align*}
                       (1-e_1) e_z  &= \lambda e_1, \nonumber \\
                      (1-e_2) e_z   &= \lambda e_2, \nonumber \\
                     e_1+e_2+e_z &=s.
\end{align*}
This yields two solutions
\begin{align}\label{eq:sole3}
                                e_1&=e_2= \left (  s+2+\lambda-\sqrt{ (s+2+\lambda)^2-8s  } \right  )/4,\\
                                e_z&=\left (  s-2-\lambda+\sqrt{ (s+2+\lambda)^2-8s  } \right  )/2,\nonumber
\end{align}
and
\begin{align*}
                                e_1&=e_2= \left (  s+2+\lambda+\sqrt{ (s+2+\lambda)^2-8s  } \right  )/4,\\
                                e_z&=\left (  s-2-\lambda-\sqrt{ (s+2+\lambda)^2-8s  } \right  )/2,
\end{align*}
It is straightforward to verify that in the latter solution~$e_z<0$, so this is not a feasible solution.
The solution~\eqref{eq:sole3} does belong to~$L_s$, so the system admits a unique equilibrium in~$L_s$.
Fig.~\ref{fig:e1e2e3} depicts trajectories of~\eqref{eq:simp_3} for three  initial
conditions in~$L_1$, namely, $\begin{bmatrix}1&0&0 \end{bmatrix}  $, $\begin{bmatrix}0&1&0 \end{bmatrix}  $,
 and~$\begin{bmatrix}0&1/2&1/2 \end{bmatrix}  $, and the equilibrium point~\eqref{eq:sole3}
for~$s=\lambda=1$. It may be seen that every one of these
  trajectories converges to~$e$.~$\square$
\end{Example}

\subsection{Contraction}
Contraction theory is a powerful tool for analyzing nonlinear dynamical systems
(see, e.g.,~\cite{LOHMILLER1998683}), with applications
to many models from systems biology~\cite{sontag_cotraction_tutorial,entrain2011,cast15}.
In a contractive system, the distance between any two trajectories
decreases at an exponential rate.
It is clear that the  RFMNP is not a contractive system on~$\Omega$, with respect to any norm,
 as it admits more than a single equilibrium point.
Nevertheless,
the next result shows that the  RFMNP is
  \emph{non-expanding} with
  respect to the~$\ell_1$ norm: $|q|_1=\sum_{i=1}^d |q_i|$.
\begin{Proposition}\label{prop:distance}
                                    For any~$a,b \in \Omega$,
                                    \be\label{eq:dist_fixed}
                                             \left   |\begin{bmatrix} x(t,a)\\z(t,a)\end{bmatrix}
																								-\begin{bmatrix} x(t,b)\\z(t,b)\end{bmatrix} \right  |_1\leq |a-b|_1,\quad \text{for all } t\geq 0.
                                    \ee
\end{Proposition}
In other words,
 the~$\ell_1$ distance between trajectories can never increase.

Pick~$a \in \Omega$, and let~$s:=1_d' a$. Substituting~$b=e_{L_s}$ in~\eqref{eq:dist_fixed}
 yields
\be\label{eq:post}
             \left   |\begin{bmatrix} x(t,a)\\z(t,a)\end{bmatrix} -e_{L_s}\right |_1\leq |a- e_{L_s} |_1,\quad \text{for all } t\geq 0.
\ee
This means that the convergence to the equilibrium point~$e_{L_s}$ is monotone in the sense that the $\ell_1$ distance to~$e_{L_s}$
can never increase.
Combining~\eqref{eq:post}
with
Theorem~\ref{thm:main}
  implies that every equilibrium point of the~RFMNP  is \emph{semistable}~\cite{Hui20082375}.


\subsection{Entrainment}

Many important biological processes
are periodic. Examples include circadian clocks and the cell cycle division process.
Proper functioning requires certain   biological systems to follow these periodic patterns, i.e. to \emph{entrain}
 to the  periodic excitation.

In the context of translation, it has been shown that both the
RFM~\cite{RFM_entrain}  and the RFMR~\cite{RFMR} entrain to periodic translation rates, i.e.
if all the transition rates   are periodic  time-varying functions, with a common (minimal) period~$T>0$
then each state variable converges to a periodic trajectory, with a period~$T$.
Here we show that the same property holds for the RFMNP.

We say that a function~$f$ is~$T$-periodic if~$f(t+T)=f(t)$ for all~$t$. Assume
that  the~$\lambda^j_i$'s in the RFMNP  are  time-varying  functions
satisfying:
\begin{itemize}
                        \item there exist~$0<\delta_1<\delta_2$ such that~$\lambda^j_i(t) \in [\delta_1,\delta_2]$ for all~$t \geq 0$ and all $j \in \{1,\dots,m\}$ ,~$i \in \{1,\dots,n_j\}$.
                        \item there exists a (minimal) $T>0$ such that all the~$\lambda^j_i$'s are~$T$-periodic.
\end{itemize}
We refer to the model in this case as the \emph{periodic ribosome flow model network with a pool}~(PRFMNP).
\begin{Theorem}  \label{thm:period}
                  Consider the PRFMNP.
                  Fix an arbitrary~$s >0$. There exists a unique function~$\phi_s:\R_+ \to \Int(\Omega) $, that is~$T$-periodic,
                   and        for any~$a\in L_s$ the solution of the PRFMNP converges to~$\phi_s$.
\end{Theorem}
In other words, every level set~$L_s$ of~$H$
contains a unique periodic solution, and every
solution of the~PRFMNP emanating from~$L_s$  converges to this solution.
Thus,  the PRFMNP  entrains (or phase locks) to the periodic excitation in the~$\lambda^j_i$'s.
This implies in particular that all the protein production rates
converge  to a periodic pattern with period~$T$.

 Note that since a constant function is a periodic function for any~$T$,
 Theorem~\ref{thm:period} implies entrainment to a
 periodic trajectory in the particular case where one of the
  $\lambda^j_i$'s oscillates, and all the other are constant.
Note also that the stability result in
Theorem~\ref{thm:main} follows from Theorem~\ref{thm:period}.

\begin{figure*}[t]
  \begin{center}
  \includegraphics[scale=.6]{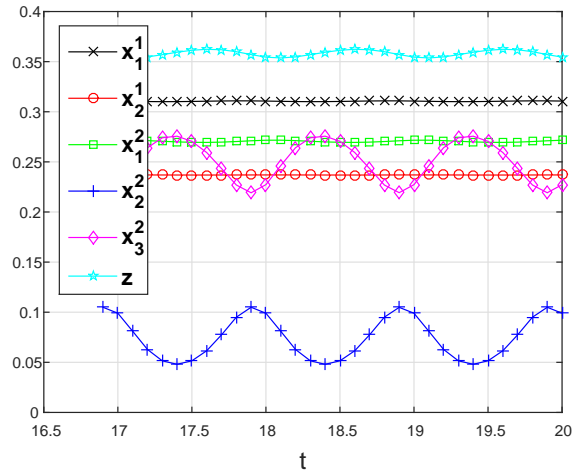}
  \caption{Trajectories of the PRFMNP in Example~\ref{exa:perio} as a function of time. }\label{fig:period}
  \end{center}
\end{figure*}

\begin{Example}\label{exa:perio}
Consider the RFMNP~\eqref{eq:exa223} with~$G_i(z)=\tanh(z)$,
 and  all  rates equal to one except for~$\lambda_{2}^2(t)=5+4\sin(2\pi t)$. In other words, there is a single time-varying periodic rate in RFMIO~$\#2$.
Note that all these rates are periodic with a common minimal period~$T=1$.
Fig.~\ref{fig:period}
depicts the solution of this PRFMNP as a function of time~$t$ for~$16.9\leq t\leq 20 $.
The initial condition is~$z(0)=x^i_j(0)=1/4$ for all~$i,j$.
It may be seen that all   the state variables converge to a periodic
 solution.  In particular, all   state variables~$x^2_i(t)$ converge to a periodic solution with (minimal) period~$T=1$,
 and so does the pool occupancy~$z(t)$.
The~$x^1_j(t)$'s also  converge  to a periodic solution, but it is not possible to tell
from the figure whether   there are small oscillations with period~$T=1$
or   the convergence is to a constant (of course,  in both cases this is a periodic solution with period~$T=1$).
\textcolor{black}{
However, it can be shown using the first two
equations in~\eqref{eq:exa223}
that if~$z(t)$ converges to a periodic solution then so do~$x^1_1(t)$ and~$x^1_2(t)$.}
Note   that the peaks in~$x^2_3(t)$ are correlated with dips in~$x^2_2(t)$, this is because when~$\lambda^2_2(t)$ is high on some time interval, i.e. the transition rate from site~$2$ to site~$3$  is high,  there is  a high flow of ribosomes
from site~$2$ to site~$3$ during this interval.~$\square$
\end{Example}

From the biophysical point of view
this means that the coupling between the mRNA molecules
can induce periodic oscillations in \emph{all}
 the protein production rates even when all the transition rates in the molecules are constant,
except for a single rate in a single molecule  that oscillates  periodically.
\textcolor{black}{
 The translation rate of codons is affected among others  by the tRNA supply (i.e. the intracellular abundance of the different tRNA species) and demand (i.e. total number of codons from each type on all the mRNA molecules) (see for example \cite{Pechmann2013}). Thus, the translation rate of a codon(s) is affected by changes in the demand (e.g. oscillations in mRNA levels) or by changes in the supply (e.g. oscillations in tRNA levels). The results reported here may suggest that oscillations in the mRNA levels of some genes or in the concentration of some tRNA species (that occur for example during the cell cycle \cite{Spellman1998,Frenkel2012}), can induce oscillations in the translation rates of the rest of the genes.
}

\subsection{Competition}
We already know that any trajectory of the RFMNP converges to an equilibrium point.
A natural question is how will a change in the parameters (that is, the transition rates) affect this equilibrium point.
For example, if we increase  some transition rate~$\lambda^j_i$ in RFMIO $\#j$,
how will this affect the steady-state production rate in the other RFMIOs?
  Without loss of generality,
we assume that the change is in a transition rate of RFMIO~$\#1$.
\begin{Theorem}\label{th:sa}
Consider an  RFMNP  with~$m$ RFMIOs with dimensions~$n_1,\dots,n_m$.
Let~$\lambda:=\begin{bmatrix} \lambda^1_0 & \dots & \lambda^m_{n_m}\end{bmatrix}'$ denote the set of all parameters of the RFMNP, and let
\[
e=\begin{bmatrix}
 e^1_1 \; \dots \; e^1_{n_1} \; e^2_1 \; \dots  \; e^2_{n_2}\; \dots \;  e^m_{1} \; \dots \; e^m_{n_m} \; e_z
 \end{bmatrix}' \in (0,1) ^{n_1+\dots +n_m} \times \R_{++}
\]
 denote the equilibrium point of the RFMNP on some fixed level set of~$H$. Pick~$i\in\{ 0,\dots,n_1 \}$.
Consider the RFMNP obtained by modifying~$\lambda^1_i$ to~$ \bar \lambda^1_i $, with~$ \bar \lambda^1_i>\lambda^1_i$.
Let~$\bar e$ denote the equilibrium point in the new RFMNP and let~$\tilde e := \bar e - e $.
Then
\begin{align}
  & \tilde e^1_i < 0,\text{ and }
   \tilde e^1_j > 0, \text{ for all } j \in \{i+1,...,n_1\} , \label{sensitivity_proof_1}\\
  & \sgn(\tilde e^i_j) = \sgn(\tilde e_z), \text{ for all } i \ne 1 \text{ and all } j .\label{sensitivity_proof_3}
\end{align} 
\end{Theorem}
(In the case $i=0$, the condition $\tilde e^1_i < 0$  is vacuous.)

Increasing~$\lambda_i^1$ means that  ribosomes flow ``more easily'' from site~$i$ to site~$i+1$ in RFMIO~$\#1$.
Eq.~\eqref{sensitivity_proof_1} means that the effect on the density in this RFMIO is
that the number of ribosomes in site~$i$ decreases,
and the number of ribosomes in  all the sites to the right of site~$i$ increases.
Eq.~\eqref{sensitivity_proof_3} describes the effect on the steady-state densities in all the
other RFMIOs and the pool:
either \emph{all} these steady-state values increase or they all decrease.
The first case agrees with the results in Example~\ref{exp:current} above.

Note that~\eqref{sensitivity_proof_1} does not provide information on the change in~$e^1_j$, $j<i$.
Our simulations show that any of these values may
either  increase or decrease, with the outcome depending  on the various  parameter values.
Thus,
 the amount of information provided by~\eqref{sensitivity_proof_1} depends on~$i$.
In particular,
	when~$\lambda^1_{n_1}$ is changed to~$\bar \lambda^1_{n_1}>\lambda^1_{n_1}$ then
	the information provided by \eqref{sensitivity_proof_1} is only that
\[
\tilde e^1_{n_1} < 0.
\]
Much more information is available when~$i=0$.
\begin{Corollary}\label{coro:sc}
Suppose that~$\lambda^1_0$ is changed to~$\bar \lambda^1_0>\lambda^1_0$.
Then
\be\label{eq:mipt}
   \tilde e^1_j > 0, \text{ for all } j \in \{1,...,n_1\} ,
\ee
and
\be \label{eq:mio2}
    \tilde e^i_j<0, \text{ for all } i \ne 1 \text{ and all } j,  \text{ and }    \tilde e_z<0.
\ee
\end{Corollary}

Indeed, for~$i=0$,~\eqref{sensitivity_proof_1} yields~\eqref{eq:mipt}. Also,
we know that the   changes in the densities in all other RFMIOs and the pool have the same sign.
This sign cannot be positive, as combining this with~\eqref{eq:mipt}
    contradicts the conservation of ribosomes, so~\eqref{eq:mio2} follows.

In other words,   increasing~$\lambda^1_0$ yields an increase in \emph{all}
 the densities in RFM~$\#1$, and a decrease in all the other densities. This makes sense, as
increasing~$\lambda^1_0$ means that it is easier  for ribosomes to bind to the mRNA molecule.
This increases  the total number of ribosomes along this molecule and, by competition,
decreases all the densities in the other molecules and the pool. Note that this special case agrees well
with the results described in~\cite{Gyorgy2015} (see~\eqref{eq:dom}).

It is important to emphasize, however, that there are various possible intracellular mechanisms that may affect $\lambda^j_i$, $i>0$. For example, synonymous mutation/changes (in endogenous or heterologous) genes inside the coding region may affect the adaptation of codons to the tRNA pool (codons that are recognized by tRNA with higher intracellular abundance usually tend to be translated more quickly~\cite{Dana2014}), the local folding of the mRNA (stronger folding tend to decrease elongation rate \cite{TullerGB2011}), or the interaction/hybridization between the ribosomal RNA and the mRNA \cite{Li2012} (there are nucleotides sub-sequence that tend to interact with the ribosomal RNA, causing  transient pausing of the ribosome, and delay the translation elongation rate). Non synonymous mutation/changes inside the coding region may also affect the elongation for example via the interaction between the nascent peptide and the exit tunnel of the ribosome \cite{Lu2008,Sabi2015}.
In addition, intracellular changes in various translation factors (e.g. tRNA levels, translation elongation factors, concentrations of amino acids, concentrations of Aminoacyl tRNA synthetase) and, as explained above, the mRNA levels can also affect elongation rates. Furthermore, various recent studies have demonstrated that manipulating the codons of a heterologous gene  tend to result in significant changes in the translation rates and protein levels of the gene \cite{Kudla2009,Welch2009,Ben-Yehezkel2015}.

Thus, our study is relevant to fundamental biological phenomena that are not covered in models that do not take into account
the elongation dynamics.  

\begin{figure*}[t]
  \begin{center}
  \includegraphics[scale=.6]{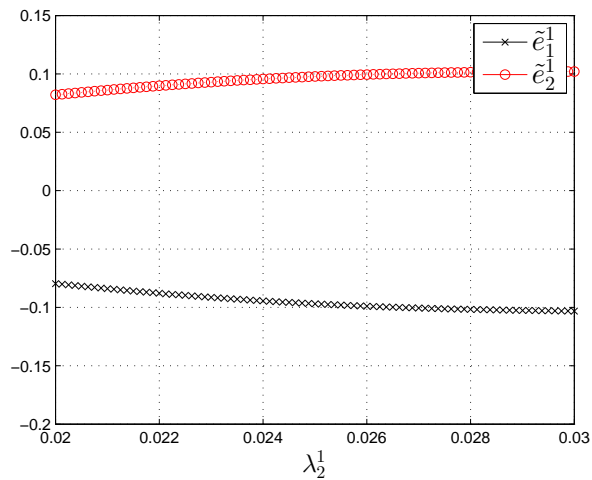}
  \caption{Steady-state perturbations in  RFM~$\#1$ in  the RFMNP in Example~\ref{exa:neg} when
	changing~$\lambda^1_1=1$ to~$\bar \lambda^1_1=10$ for various values of~$\lambda_2^1$. }\label{fig:neg1}
  \end{center}
\end{figure*}
\begin{figure*}[t]
  \begin{center}
  \includegraphics[scale=.6]{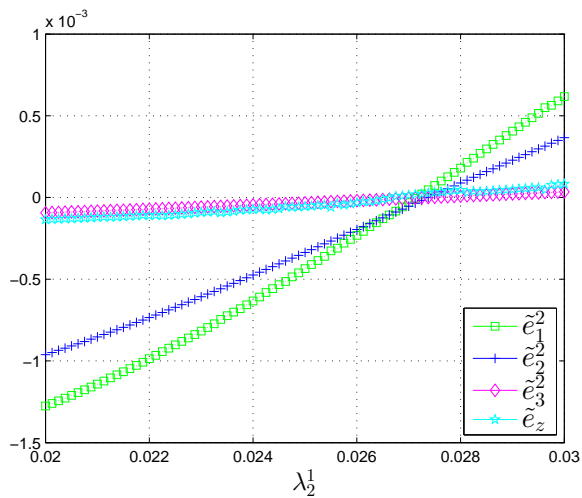}
  \caption{Steady-state perturbations in   RFM~$\#2$ in  the RFMNP in Example~\ref{exa:neg}  when
	changing~$\lambda^1_1=1$ to~$\bar \lambda^1_1=10$ for various values of~$\lambda_2^1$. }\label{fig:neg2}
  \end{center}
\end{figure*}

\begin{Example}\label{exa:neg}
Consider the RFMNP in~\eqref{eq:exa223} with~$G_i(z)=z$, $\lambda^1_0=\lambda^2_0=1$,
$\lambda^2_1=\lambda^2_2=0.1$, $\lambda^2_3=1$, and initial condition~$(1/4)1_6$.
We consider a range of values for~$\lambda^1_2$. For each fixed value, we simulated the dynamics until
steady-state for two cases:~$\lambda^1_1=1$ and~$\bar \lambda_1^1=10$.
Fig.~\ref{fig:neg1} depicts~$\tilde e^1_1,\tilde e^1_2$ for the various fixed values of~$\lambda^1_2$.
It may be seen that we always have~$\tilde e^1_1<0$ and~$\tilde e^1_2>0$.
Fig.~\ref{fig:neg2} depicts~$\tilde e^2_i$, $i=1,2,3$, and~$\tilde e_z$ for the various fixed values of~$\lambda^1_2$.
It may be seen that for a small value of~$\lambda^1_2$ all the~$\tilde e^2_i $'s
and~$\tilde e_z$ are negative, whereas for  large values of~$\lambda^1_2$ they all become positive.~$\square$
\end{Example}


Theorem~\ref{th:sa} implies   that when
 the codons of a gene are modified into ``faster codons''
(either via synthetic
engineering or during evolution)  then either
all the translation rates of the other genes increase, or they all decrease.
However, Theorem~\ref{th:sa} does not provide information on when
  each of these two cases happens.
In order to address this, we need to calculate  derivatives
of the equilibrium point coordinates with respect to the rates.
The next result shows that these derivatives are well-defined.
Denote the mapping from the parameters to the unique  equilibrium point in~$\Int(\Omega)$ by~$\alpha $, that is, $e^i_j=\alpha^i_j(\lambda,H(0))$, $i=1,\dots,m$, $j=1,\dots, n_i$.
\begin{Proposition}\label{prop:different}
The derivative~$\frac{\partial } {\partial \lambda^p_q}\alpha^i_j(\lambda,H(0))$ exists for all~$i,j,p,q$.
\end{Proposition}

The next  example uses these derivatives to obtain
information on the two cases that can take place as we change one of the rates.
\begin{Example}
Consider an RFMNP with~$m=2$ RFMIO's with lengths~$n$ and~$\ell$.
 To simplify the notation, let~$e=\begin{bmatrix}e_1,\dots,e_{n}\end{bmatrix}'$
[$v=\begin{bmatrix}v_1,\dots,e_{\ell}\end{bmatrix}'$]
denote the equilibrium point of RFMIO~$\#1$ [RFMIO~$\#2$], and
let~$\lambda_i$, $i=0,\dots,n$, denote the
  rates along RFMIO~$\#1$.
Suppose that~$\lambda_1$ is changed to~$\bar \lambda_1$.
  Differentiating  the steady-state equations
\begin{align}
  \lambda_0 G_1(e_z)(1-e_1)=\lambda_1 e_1(1-e_2) &= ... = \lambda_{n-1} e_{n-1}(1-e_n) = \lambda_n e_n, 	\nonumber  \\
  \sum_{i=1}^n e_i+\sum_{j=1}^\ell v_j+e_z&=H(0) 	,				    	\nonumber
\end{align}
w.r.t.~$\lambda_1$ yields
\begin{align*}
  \lambda_0  G_1'(e_z) e_z'(1-e_1)- \lambda_0  G_1(e_z)e_1'&= \lambda_n e_n'  , \\
  e_z' +\sum_{j=1}^\ell v_j' &=-\sum_{i=1}^n e_i'  ,
\end{align*}
where we use the notation~$f':=\frac{\partial}{\partial \lambda_1}f$. These two equations yield
\begin{align*}
  (\lambda_n + \lambda_0  G_1'(e_z)(1-e_1) ) e_z'+\lambda_n\sum_{j=1}^{\ell} v_j'&= ( \lambda_0 G_1(e_z)-\lambda_n
	) e_1'-\lambda_n \sum_{i=2}^{n-1} e_i'.
\end{align*}
Recall that~$G_1(e_z)>0$, $G_1'(e_z)>0$, $ \lambda_j>0$ for all~$j$,  and~$ 0<e_p<1$ for all $p$. Also, by
 Theorem~\ref{th:sa},
$e_1'<0$,~$e_j'>0$, for all $j \in {2,...,n}$,
and~$\sgn(e_z')=\sgn(v_k')$, for all $k$. Thus,
\begin{align}\label{eq:diff:conc}
  \sgn(e_z')=\sgn(v_j')&= \sgn \left( ( \lambda_0  G_1(e_z)-\lambda_n) e_1'-\lambda_n \sum_{i=2}^{n-1} e_i' \right ).
	\end{align}
This means that the sign of the
change in the densities in all the other RFMIO's
and   the pool depends on several steady-state quantities including terms
related to the
initiation  rate~$\lambda_0 G_1(e_z)$ and exit rate $\lambda_n$ in RFMIO~$\#1$, and also  the change in the total density~$ \sum_{i=1}^{n-1} e_i'$
in this RFMIO.

In the particular  case~$n=2$ (i.e., a very short RFM), Eq.~\eqref{eq:diff:conc}  becomes:
\begin{align}\label{eq:diff:conc2}
  \sgn(e_z')=\sgn(v_j')&= \sgn(  \lambda_2 -\lambda_0 G_1(e_z)  ).
\end{align}
Note that~$\lambda_0 G_1(e_z)$ [$\lambda_2$] is the steady-state
initiation [exit] rate in RFMIO~$\#1$.
Thus, $\lambda_2 -\lambda_0 G_1(e_z) >0 $ means that it is ``easier'' for ribosomes  to exit
 than to enter RFMIO~$\#1$, and in this case
\eqref{eq:diff:conc2} means that when~$\lambda_1$ is increased
the change in all other densities will be positive. This is intuitive, as more ribosomes will exit
the modified molecule and this will improve the production rates in the other molecules.
On the other-hand, if  $\lambda_2 -\lambda_0 G_1(e_z)< 0 $ then
it is ``easier'' for ribosomes  to enter
 than to exit  RFMIO~$\#1$, so increasing~$\lambda_1$ will lead
to an increased  number of ribosomes in RFMIO~$\#1$ and, by competition,
to a decrease in the production rate in all the other RFMIOs.~$\square$
\end{Example}

Note that in the example above increasing~$\lambda_1$ always
  increases   the
steady-state production rate~$R=\lambda_2 e_2$  in RFM~$\#1$ (recall that $e_2'>0$).
One may expect that this will always lead to an increase in the production rate
in the second RFMIO as well.
However,
the behavior in the RFMNP is more complicated because the  shared  pool
generates  a  feedback connection between the
RFMIO's in the network. In particular, the effect on the other RFMIO's
depends not only on the modified  production rate of RFMIO~$\#1$, but on other factors including
     the change in the \emph{total} ribosome density in RFMIO~$\#1$ (see~\eqref{eq:diff:conc}).

This analysis of a very short RFM suggests that the steady-state
initiation rate  of the mRNA with the modified codon plays an important role in determining
 the effect of modifications in the network. If this initiation rate is relatively
low (so it becomes the rate limiting factor), as believed to be the case in most    endogenous genes~\cite{Jacques1990},  then the
 increase in the rate of one codon of the mRNA  increases the translation rate in all the other
mRNAs, whereas when this initiation rate is high then the opposite effect is obtained.
This latter case  may  occur for example when a  heterologous gene is highly expressed and thus  ``consumes''
 some of  the available elongation/termination factors   making the elongation rates the rate limiting factors.

\section{Discussion}
We introduced  a new model, the RFMNP, for large-scale simultaneous translation and competition for ribosomes
that combines several  RFMIOs interconnected via a dynamic pool of free ribosomes.
To the best of our knowledge, this is the first model of a network composed of interconnected RFMIOs.
The  RFMNP  is amenable to analysis because it is a monotone dynamical system
that admits a non-trivial first integral.
 Our results show that the RFMNP  has several nice properties:   it is an irreducible
   cooperative dynamical system admitting a continuum of linearly ordered
equilibrium points, and
every trajectory converges to an equilibrium point. The RFMNP
 is  also on the ``verge of contraction'' with respect to the~$\ell_1$ norm,
and it entrains
  to periodic transition rates with a common period.
	The fact that the total number of ribosomes in the network
		is conserved means that local
	 properties of any mRNA molecule (e.g.,  the abundance of the corresponding  tRNA molecules)
	  affects  its own translation rate, and via competition, also
	   \emph{globally} affects    the translation rates of all
		the other mRNAs in the network.

An important implication of our analysis and simulation results is that there are regimes and
parameter values where there is a strong
coupling
 between the different ``translation components'' (ribosomes and mRNAs) in the cell.
Such regimes  cannot be studied using models for translation of a single isolated mRNA molecule.
The RFMNP  is specifically important when studying   highly expressed genes with many mRNA molecules and ribosomes
 translating them because
 the dynamics of such genes   strongly  affects   the  ribosomal pool.
For example, changes in the translation dynamics of a  heterologous gene which is expressed with a very strong promoter,
 resulting in very high mRNA copy number should affect the entire tRNA pool,
 and thus the translation of other endogenous genes.
Highly expressed endogenous genes "consume" many ribosomes. Thus,
 a mutation that affects  their (``local'') translation rate is expected to affect also the translation dynamics of other mRNA molecules.
Studying the evolution of such genes  should be based on understanding the global
effect of such mutations
 using a computational model such as the~RFMNP.

On the other hand, we can approximate the  dynamics
of  genes that are not   highly expressed  (e.g., a gene  with mRNA levels that are $0.01\%$ of the mRNA levels in the cell)
   using a 
 single RFM. In  this case, the relative effect of the mRNA  on all other mRNAs 
 is expected to be limited.

Our analysis shows  that increasing the translation initiation rate of a heterologous gene will always have a negative effect on the translation rate of other genes (i.e. their translation rates decrease) and vice versa. The effect of increasing 
[decreasing] the translation rate of a codon of the heterologous gene on the translation rate of other genes is more complicated: while it always increases [decreases]
 the translation rate of the heterologous gene it may either increase or decrease the translation rate of all other genes. The specific outcome of such a manipulation can be studied  using the RFMNP  with  the parameters   based on the heterologous genes and the host genome.


Our analysis suggests that the effect of improving the transition
 rate of a codon in an mRNA molecule on the production  rate of other genes and the pool of ribosomes   
depends on the initiation rate in the modified mRNA. 
 When the initiation rate is very low the effect is expected  to be positive (all other production
 rates increase). However, if the initiation rate is high   the effect may be negative. This
 may partially explain  the selection for slower codons in highly expressed genes that  practically decrease the initiation rate~\cite{Tuller2015,Tuller2010c}.
This relation may also suggest a new factor that contributes to the evolution of highly expressed genes towards higher elongation and termination rates (i.e., the tendency of highly expressed genes to include  ``fast'' codons).  Indeed,   lower  elongation rates (and thus a relatively high initiation rate) may decrease the production rates of other mRNAs that are needed for proper functioning of the organism.


The RFM, and thus also the model described here,  do  not capture certain
 aspects of mRNA translation. For example,   eukaryotic ribosomes may translate mRNAs in multiple cycles before entering the free ribosomal pool \cite{haar_survey,RFM_feedback,Kopeina2008}.
This phenomenon may perhaps  be modeled by adding positive feedback~\cite{RFM_feedback} in  the RFMNP.
In addition,  different genes are transcribed at different rates, resulting in a
different number of (identical) mRNA copies for different genes. This can be modeled using
 a set of identical RFMs for each gene. Such a model can help in understanding   how changes in mRNA levels of one gene affect the translation rates of all the mRNAs. The analysis
   here suggests that modifying
 the mRNA levels of a gene will affect  the translation rates of all other   genes in the same way.
These and other aspects of biological translation
may be integrated in our model in future studies.
Ref.~\cite{mitiga_vecc} develops the notion of  the realizable region for steady-state
gene expression under resource limitations, and methods
 for mitigating the effects of ribosome competition. Another  interesting
 research direction is studying these topics in the context
 of the RFMNP.

The results reported here can be studied experimentally by designing and expressing a library of heterologous genes \cite{Welch2009,Ben-Yehezkel2015,Kudla2009}. The effect of the manipulation of a codon  (i.e, increasing or decreasing its rate) of the heterologous gene on the ribosomal densities and translation rates of all the mRNAs (endogenous and heterologous) can be performed  via ribosome profiling \cite{Ingolia2009} in addition to measurements of mRNA levels, translation rates, and protein levels~\cite{Schwanhausser2011}.

 We  believe that   networks of interconnected RFMIOs may prove to be powerful modeling  and analysis tools
 for other natural and artificial systems as well. These include
 communication networks, intracellular trafficking in the cell,
 coordination of large groups of organisms (e.g., ants),   traffic control, and more.


\section*{Acknowledgments}
We thank Yoram Zarai for helpful comments.  


\section*{Appendix: Proofs}

{\sl Proof of Proposition~\ref{prop:inv}.}
We require the following result on repelling  boundaries and persistence.
\begin{Lemma}\label{lem:per}
Consider a time-varying system
\be \label{eq:gensys}
\dot x = f(t,x)
\ee
whose trajectories evolve    on  $\Omega: = I_1\times I_2\times \ldots \times I_n\subseteq \R^n_{+}$, where each
$I_j$ is an interval of the form $[0,A]$, $A>0$, or $[0,\infty )$.  Suppose
that the time-dependent vector field $f=\begin{bmatrix}
f_1,\ldots ,f_n\end{bmatrix}'$ has the following two properties:
\bi
\item the \emph{cyclic boundary-repelling} property~{\bf (CBR)}:
For each $\delta >0$ and each sufficiently small $\Delta >0$, there exists
$K=K(\delta ,\Delta )>0$ such that, for each $k=1,\ldots ,n$ and each $t\geq 0$,
the condition
\be\label{eq:fk}
x_k (t) \leq \Delta,   \text{ and } x_{k-1}(t)\geq \delta
\ee
(all indexes  are modulo~$n$)
implies that
\be\label{eq:fkk}
f_{k }(t,x) \geq  K ;
\ee
\item for any~$i\in\{1,\dots,n\}$, and any~$s\geq 0$, $x_i(s)>0$ implies that
\be\label{eq:stay}
x_i(t)>0,\quad\text{for all } t\geq s.
\ee
\ei
Then  given any $\tau>0$ there exists   $\varepsilon  = \varepsilon (\tau)>0$,
 with~$\varepsilon(\tau) \to 0$ as~$\tau \to 0$, such that every
solution~$x(t,a)$, with~$a\not =  0$,  satisfies
\[
x_i(t,a)\geq \varepsilon   \text{ for
all }     i\in \{1,\ldots ,n\}     \text{ and all }    t\geq  \tau.
\]
\end{Lemma}
In other words, the conclusion is that after an arbitrarily short time
every~$x_i(t,a )$ is separated away from zero.

{\sl Proof of Lemma~\ref{lem:per}.}
Pick   $\tau>0$ and~$a\not =0$.
Then there exists~$i\in\{1,\dots,n\}$ such that~$a_i>0$. Since the ({\bf CBR})
condition is cyclic, we may assume
without loss of generality  that~$x_n(0)>0$.
Then~\eqref{eq:stay} implies that
there exists~$\delta>0$ such that~$x_n(t)\geq\delta$ for all~$t\in[ 0 ,\tau/n]$.

Fix~$\Delta>0$ such that {\bf (CBR)}  holds.  Let $K=K(\delta ,\Delta )$,   and define $\varepsilon _1 :=\min\{\Delta ,K\tau /n\}$.
Let $t_0\in [0,\tau/n]$ be such that $x_1(t_0)\geq \varepsilon _1$.
Such a $t_0$ exists, since by property~{\bf (CBR)},
$x_1(t)<\varepsilon _1\leq \Delta $ for all $t\in [0,\tau/n]$ would imply that
$\dot x_1(t)=f_1(t,x(t))\geq K>0$ for all $t\in [0 ,\tau/n]$, which in turn implies
$x_1(\tau/n)\geq x_1(0)+K\tau/n   \geq K\tau/n  \geq \varepsilon _1$, contradicting $x_1(\tau/n)<\varepsilon _1$.
We claim that also $x_1(t)\geq \varepsilon _1$ for every $t\geq t_0$.
Indeed, suppose otherwise.  Then, there is some $t_1>t_0$ such that
$\xi : =x_1(t_1)<\varepsilon _1$.
Let
\[
\sigma  :=\min\{t\geq t_0 \st x_1(t)\leq \xi \}>t_0.
\]
As $x_1(\sigma )\leq \xi <\varepsilon _1\leq \Delta $, and~$x_n(\sigma)>0$ property {\bf (CBR)} says that
$\dot x_1(\sigma )=f_1(\sigma ,x(\sigma )) > 0 $, so it follows that $\dot x_1(t)>0$ on an
interval $[\sigma -c ,\sigma ]$, for some $c >0$.
But then $x_1(\sigma -c ) < x_1(\sigma ) \leq  \xi $, which contradicts the minimality of
$\sigma $.
Thus $x_1(t)\geq \varepsilon _1$ for all $t\geq t_0$, and in particular
\be\label{eq:propq}
						x_1(t)\geq \varepsilon _1,\quad \text{for all } t\geq \tau/n.
\ee


 Let $\bar K:=K(\varepsilon_1 ,\Delta )$,   and define $\varepsilon_2 :=\min\{\Delta ,\bar K\tau /n\}$.
Let $t_0\in [\tau/n,2\tau/n]$ be such that $x_2(t_0)\geq \varepsilon _2$.
Such a $t_0$ exists, since by property~{\bf (CBR)} and~\eqref{eq:propq},
$x_2(t)<\varepsilon _2\leq \Delta $ for all $t\in [\tau/n , 2 \tau/n]$ would imply that
$\dot x_2(t)=f_2(t,x(t))\geq \bar K>0$ for all $t\in [\tau/n ,2 \tau/n]$, which in turn implies
$x_2(2\tau/n)\geq x_2(\tau/n)+\bar K \tau/n   \geq  \bar K\tau/n  \geq \varepsilon _2$, contradicting $x_2(2\tau/n)<\varepsilon _2$.
We claim that also $x_2(t)\geq \varepsilon _2$ for every $t\geq t_0$.
Indeed, suppose otherwise.  Then, there is some $t_1>t_0$ such that
$\xi : =x_2(t_1)<\varepsilon_2$.
Let
\[
\sigma  :=\min\{t\geq t_0 \st x_2(t)\leq \xi \}>t_0.
\]
As $x_2(\sigma )\leq \xi <\varepsilon _2\leq \Delta $, and~$x_1(\sigma) \geq \varepsilon_1$ property {\bf (CBR)} says that
$\dot x_2(\sigma )=f_2(\sigma ,x(\sigma )) > 0 $, so it follows that $\dot x_2(t)>0$ on an
interval $[\sigma -c ,\sigma ]$, for some $c >0$.
But then $x_2(\sigma -c ) < x_2(\sigma ) \leq  \xi $, which contradicts the minimality of
$\sigma $.
Thus $x_2(t)\geq \varepsilon _2$ for all $t\geq t_0$, and in particular
\[
						x_2(t)\geq \varepsilon _2,\quad \text{for all } t\geq 2\tau/n.
\]
Continuing in this manner we have that for every~$i\in\{1,\dots,n\}$ there exists~$\varepsilon_i>0$ such that
\[
					x_i(t)\geq \varepsilon_i,\quad\text{for all } t\geq i \tau/n.
\]
Thus,
\[
					x_i(t)\geq \min
					\{\varepsilon_1,\dots,\varepsilon_n\} ,\quad \text{for all } t \geq   \tau  \text{ and all }  i=1,\dots,n,
\]
and this completes the proof of Lemma~\ref{lem:per}.~$\square$

We can now  prove Proposition~\ref{prop:inv}. Consider the
  case~$m=1$, i.e. the RFMNP is given by
\begin{align}\label{eq:rf1}
                  \dot{x}_1&=\lambda_0  (1-x_1)G(x_{n+1}) -\lambda_1 x_1(1-x_2), \nonumber \\
                    \dot{x}_2&=\lambda_{1} x_{1} (1-x_{2}) -\lambda_{2} x_{2} (1-x_3) , \nonumber \\
                             &\vdots \nonumber \\
                    \dot{x}_{n-1}&=\lambda_{n-2} x_{n-2} (1-x_{n-1}) -\lambda_{n-1} x_{n-1} (1-x_n), \nonumber \\
                    \dot{x}_n&=\lambda_{n-1}x_{n-1} (1-x_n) -\lambda_n x_n  ,\nonumber\\
									\dot x_{n+1}&=\lambda_n x_n-\lambda_0(1-x_1)G(x_{n+1}) ,
\end{align}
where we write~$x_{n+1}$  instead of~$z$.
The proof in the case where~$m>1$  is similar.
We begin by showing that~\eqref{eq:rf1}
satisfies the properties in Lemma~\ref{lem:per} on~$ \Omega=[0,1]^n\times[0,\infty)$.
Fix an arbitrary~$\delta>0$.
If~$x_1 \leq \Delta$ and~$x_{n+1}\geq\delta $  then
\begin{align*}
        f_1&= \lambda_0(1-x_1)G(x_{n+1})-\lambda_1x_1(1-x_2) \\
           &\geq K_1 ,
\end{align*}
where~$K_1: =\lambda_0( 1-\Delta)G(\delta) -\lambda_1 \Delta   $.
Now pick~$1<k<n$. If~$x_k\leq\Delta$ and
 $x_{k-1}\geq \delta$   then
\begin{align*}
        f_k&=\lambda_{k-1} x_{k-1} (1-x_k)-\lambda_k x_k(1-x_{k+1}) \\
           &\geq {K}_k,
\end{align*}
where
\[
  {K}_k:=  \lambda_{k-1} \delta(1-\Delta)- \lambda_k \Delta .
  \]
If~$x_n\leq\Delta$ and
 $x_{n-1}\geq \delta$ for~$1\leq i\leq n-1$ then
\begin{align*}
        f_n&=\lambda_{n-1} x_{n-1} (1-x_n)-\lambda_n x_n  \\
           &\geq {K_{n-1}}.
\end{align*}
Finally, if~$x_{n+1} \leq\Delta$ and
 $x_{n}\geq \delta$ then
\begin{align*}
        f_{n+1}&= \lambda_n x_n-\lambda_0(1-x_1)G(x_{n+1})   \\
           &\geq {K_{n+1}},
\end{align*}
where~$K_{n+1}:=\lambda_n \delta -\lambda_0 G(\Delta) $.
Thus,~\eqref{eq:fkk} holds for~$K:=\min \{K_1,\dots,K_{n+1}\}$,
and clearly~$K>0$ for all~$\Delta>0$ sufficiency small. Thus,~\eqref{eq:rf1}
  satisfies {\bf (CBR)}.
	
To show that~\eqref{eq:rf1} also satisfies~\eqref{eq:stay}, note that for all~$k\in\{1,\dots,n\}$ and all~$x\in\Omega$,
$\dot x_k\geq -\lambda_k x_k$, and
that by the properties of~$G(\cdot)$,~$\dot x_{n+1} \geq -s x_{n+1}$	for all~$x_{n+1}>0$ sufficiently small. Thus,~\eqref{eq:rf1} satisfies all
the conditions in Lemma~\ref{lem:per} and this implies that  for any~$\tau>0$
there exists~$\varepsilon=\varepsilon(\tau/2)>0$
such that
\be\label{eq:wwkm}
x_i(t,a)\geq\varepsilon,\quad \text{for all } i\in\{1,\dots,n+1\} \text{ and all }t\geq \tau/2.
	\ee
This proves the first part of 	 Proposition~\ref{prop:inv}. To complete the proof,
define
\be \label{eq:deffy}
y_i(t):=1-x_{n +1-i}(t) , \quad i=1,\ldots,n,
\ee
  and~$y_{n+1}(t):=x_{n+1}(t)$.
Then
\begin{align}\label{eq:rf1y}
                \dot y_1&= \lambda_n (1-y_1) - \lambda_{n-1} y_1   (1-y_2) ,\nonumber\\
							  \dot y_2&= \lambda_{n-1}y_1 (1-y_2) - \lambda_{n-2} y_2 (1-y_{3}) ,\nonumber\\
									\vdots \nonumber\\
									\dot y_{n-1}& = \lambda_{2}y_{n-2} (1- y_{n-1})  - \lambda_{1} y_{n-1}(1-y_n)     ,\nonumber \\
									\dot y_{n}& =   \lambda_1 y_{n-1}(1-y_n) - \lambda_0   G(y_{n+1})  y_n,\nonumber \\
									\dot y_{n+1}& =   \lambda_n (1-y_1) -\lambda_0   G(y_{n+1})y_n .\nonumber
\end{align}
The first~$n$ equations here are an RFM with a time-varying
 exit rate~$\lambda_0 G(y_{n+1}(t))$.
We already know that~$y_{n+1}(t)\geq\varepsilon$ for all~$t\geq\tau/2$ and the results in~\cite{RFM_entrain}
imply that there exists~$\varepsilon_1=\varepsilon_1(\tau )>0$ such that
\[
y_i(t,a)\geq\varepsilon_1,\quad \text{for all } i\in\{1,\dots,n \} \text{ and all }t\geq \tau .
	\]
Combining  this with~\eqref{eq:deffy} completes the proof of Prop.~\ref{prop:inv}.~$\square$

\noindent {\sl Proof of Proposition~\ref{prop:mono}.}
 The
  Jacobian matrix~$J $ of the RFMNP
  has the form
\begin{equation}\label{eq:jacobian}
 J =\begin{bmatrix}
 J_{1,1}  & 0 & ... & 0 & J_{1,{m+1}}\\
 0 & J_{2,2} & ... & 0 & J_{2,{m+1}}\\
 \vdots &  \vdots & ... &\vdots \\
 0 & 0 & ... & J_{m,m} & J_{m,{m+1}}\\
 J_{{m+1},1} & J_{{m+1},2} & ... & J_{{m+1},m} & J_{{m+1},{m+1}}
\end{bmatrix} \in \mathbb{R}^{d \times d}.
\end{equation}
Here~$J_{i,i}\in \R^{n_i \times n_i}$, $i=1,\dots,m$, is the Jacobian of RFMIO~$\#i$ given by
\begingroup\makeatletter\def\f@size{9}\check@mathfonts
\begin{equation}
\begin{bmatrix}
-\lambda^i_0 G_i(z) - \lambda^i_1 (1-x^i_2) & \lambda^i_1 x^i_1                                 & 0            &           &  0&     0\\
\lambda^i_1(1-x^i_2)                   & -\lambda^i_1 x^i_1-\lambda^i_2(1-x^i_3)               & \lambda^i_2 x^i_2     &        & 0   & 0\\
0                                  &  \lambda^i_2  (1-x^i_3)                           & -\lambda^i_2 x^i_2 -\lambda^i_3(1-x^i_4)  &   & 0 &0\\
                                    &                                              &     \vdots \\
0                                   &                                             0 &      0          &                  & -\lambda^i_{n_i-2}x^i_{n_i-2} -\lambda^i_{n_i-1} (1-x^i_{n_i})  & \lambda^i_{n_i-1}x^i_{n_i-1} \\
0                                       &                   0 & 0                  &            & \lambda^i_{n_i-1}(1-x^i_{n_i}) & -\lambda^i_{{n_i}-1}x^i_{{n_i}-1} -\lambda^i_{n_i} \nonumber
\end{bmatrix},
\end{equation}
  \endgroup
 and the other blocks are
\begin{align*}
 J_{{m+1},i}&= [\lambda^i_0 G_i(z) \ 0 \ ... \ 0 \ \lambda^i_{n_i}]  \in \mathbb{R}^{1 \times n_i} , \\
 J_{i,{m+1}}&= [\lambda^i_0 (1-x_1^i)G_i'(z) \ 0 \ ...   \ 0]'  \in \mathbb{R}^{n_i \times 1}  ,
\end{align*}
$i =1,\dots,m$, and
\begin{equation}
 J_{{m+1},{m+1}}= - \sum_{j=1}^m \lambda^j_0 (1-x_1^j)G_j'(z)  \in \mathbb{R}. \nonumber
\end{equation}
Since~$x_i \in[0,1]$ and~$\lambda^i_j>0$, every off-diagonal entry of~$J_{i,i}$ is non-negative.
Since~$z\geq 0$, every entry of~$ J_{{m+1},j}, J_{i,{m+1}}$, $i=1,\dots,m$, is also nonnegative. We conclude that
every non-diagonal entry of~$J$ is non-negative, and this implies~\eqref{eq:abab} (see~\cite{hlsmith}).
To prove~\eqref{eq:strongabab}, note that for any point in~$ \Int(\Omega)$ (i.e.,~$x_j \in (0,1)$, $j=1,\dots,d-1$, and~$z>0$)
every entry on the super- and sub-diagonal of~$J_{i,i}$ is strictly positive. Also,~$G_i(z)>0$, $i=1,\dots,m$.
This implies that the matrix~$J$ in~\eqref{eq:jacobian} is irreducible. Combining this with Prop.~\ref{prop:inv}
completes the proof.~$\square$

\noindent {\sl Proof of Theorem~\ref{thm:main}.}
Since the RFMNP is  a cooperative irreducible system on~$\Int(\Omega)$ with a non-trivial   first integral,
    Thm.~\ref{thm:main} follows from combining Prop.~\ref{prop:inv} with the results in~\cite{mono_plus_int} (see also~\cite{RFMR},~\cite{Mierc1991}
    and~\cite{mono_chem_2007}
 for   related ideas).~$\square$

\noindent {\sl Proof of Proposition~\ref{prop:distance}.}
Recall that the matrix measure~$\mu_1(\cdot):\R^{n \times n} \to \R$ induced by the~$\ell_1$ norm
is
\[
            \mu_1(A)=\max \{c_1(A),\dots, c_n(A)\},
\]
where~$c_i(A):=a_{ii}+\sum_{k\not = i} | a_{ki} | $,
i.e.   the sum of entries in column~$i$ of~$A$, with the off-diagonal entries taken with absolute value~\cite{vid}.
For the Jacobian of the RFMNP~\eqref{eq:jacobian}, we have~$c_i(J(x))=0$  for all~$i$ and all~$x \in \Omega$, so~$\mu_1(J(x))=0$.
Now~\eqref{eq:dist_fixed} follows from standard results in contraction theory (see, e.g.,~\cite{entrain2011}).~$\square$

\noindent {\sl Proof of Theorem~\ref{thm:period}.}
Write the PRFMNP  as~$\dot x= f(t,x)$. Then~$f(t,y)=f(t+T,y)$ for all~$t$ and~$y$.
Furthermore,~$H$ in~\eqref{eq:nrivi} is a non trivial first integral of the dynamics.
Now Theorem~\ref{thm:period} follows from the results in~\cite{mono_periodic} (see also~\cite{mono_periodic_96}).
The fact that~$\phi_s \in \Int(\Omega)$ follows from Proposition~\ref{prop:inv}.~$\square$

\noindent {\sl Proof  of Theorem~\ref{th:sa}.}
To simplify the presentation,
we prove the theorem for the case~$m=2$ and change some of the  notation. The proof in the general case is very similar.
Let~$n$ [$\ell$] denote the dimension of the first [second] RFMIO,
and let~$\lambda_i$ [$\zeta_i$] denote the   rates in the first [second] RFMIO.
We denote the state-variables of RFMIO~$\#1$ by~$x_i$, $i=1,\dots,n $,
 and those of the second RFMIO by~$y_i$, $i=1,\dots,\ell$.
Let~$e_i$, $i=1,\dots,n$, [$v_j$, $j=1,\dots,\ell$] denote the equilibrium point of the first [second] RFMIO.
Then  we need to prove that
\begin{align}
  &  \tilde e_i < 0,\text{ and }
   \tilde e_j > 0, \text{ for all } j \in \{i+1,...,n \} , \label{eq:need1}\\
  & \sgn(\tilde v_1) = \dots = \sgn(\tilde v_\ell) =\sgn(\tilde e_z), \label{eq:need2}
\end{align}
where~$\tilde v_j:=\bar v_j-v_j$.
At  steady state,  the RFMNP equations yield:
\begin{align}
                 \lambda_0   G_1(e_z) (1-e_1) &= \lambda_1 e_1(1-e_2), \nonumber \\
                     &= \lambda_{2} e_{2} (1-e_3) , \nonumber \\
                  &= \lambda_{3} e_{3} (1-e_4) , \nonumber \\
                             &\vdots \nonumber \\
                     &=\lambda_{n-1}e_{n-1} (1-e_n)\nonumber \\
                     &=  \lambda_n e_n  , \label{eq:eqeis} \\
                      \zeta_0 G_2(e_z) (1-v_1) &=\zeta_1 v_1(1-v_2), \nonumber \\
                    &=\zeta_{2} v_{2} (1-v_3) , \nonumber \\
                    &=\zeta_{3} v_{3} (1-v_4) , \nonumber \\
                             &\vdots \nonumber \\
                     &=\zeta_{\ell-1}v_{\ell-1} (1-v_\ell) \nonumber\\
                     &=\zeta_\ell v_\ell ,  \label{eq:eqvvis}  \\ 		
                 \lambda_n e_n +\zeta_\ell v_\ell &= \lambda_0
								G_1(e_z) (1-e_1) +\zeta_0 G_2(e_z)(1-v_1) . \label{eq:eqpool}
\end{align}
Also, since~$H$ is a first integral
\be\label{eq:fiz}
                        \sum_{i=1}^n e_i+\sum_{j=1}^\ell v_j+e_z=H(0).
\ee
Pick~$i\in\{1,\dots, n-1\}$. Consider a new RFMNP obtained by modifying~$\lambda_i$ in RFMIO~$\#1$
to~$\bar \lambda_i$, with~$\bar \lambda_i>\lambda_i$.
Then the equations for the modified equilibrium point are:
\begin{align}
                 \lambda_0   G_1(\bar  e_z) (1-\bar e_1) &=    \lambda_1 \bar  e_1(1-\bar  e_2), \nonumber \\
                     &= \lambda_{2} \bar  e_{2} (1-\bar  e_3) , \nonumber \\
                             &\vdots \nonumber \\
									&= \bar \lambda_{i} \bar  e_{i} (1-\bar  e_{i+1}) , \nonumber \\
                  &\vdots \nonumber \\					
                     &=\lambda_{n-1}\bar  e_{n-1} (1-\bar  e_n)\nonumber \\
                     &=  \lambda_n \bar  e_n  , \label{eq:eqeiswithbar} \\
                    \zeta_0  G_2(\bar e_z) (1-\bar  v_1) &=\zeta_1 \bar  v_1(1-\bar v_2), \nonumber \\
                    &=\zeta_{2} \bar  v_{2} (1-\bar v_3) , \nonumber \\
                             &\vdots \nonumber \\
                     &=\zeta_{\ell-1}\bar v_{\ell-1} (1-\bar v_\ell) \nonumber\\
                     &=\zeta_\ell \bar  v_\ell ,  \label{eq:eqvviseithbar}  \\
                 \lambda_n \bar  e_n +\zeta_\ell \bar  v_\ell &=\lambda_0 G_1(\bar  e_z) (1-\bar  e_1) +\zeta_0 G_2(\bar  e_z)(1-\bar  v_1) \label{eq:lastwithbar}.
\end{align}
Since the initial condition remains the same,
\[
                \sum_{i=1}^n   e_i+\sum_{j=1}^\ell   v_j+ e_z
								= \sum_{i=1}^n \bar e_i+\sum_{j=1}^\ell \bar v_j+\bar e_z=H(0),
\]
so
\be\label{eq:fiz:totalh}
                        \sum_{i=1}^n \tilde e_i+\sum_{j=1}^\ell \tilde v_j+\tilde e_z=0.
\ee

The last equality in \eqref{eq:eqvvis} yields
\[
 \zeta_{\ell-1}v_{\ell-1} = \frac{\zeta_\ell v_\ell}{1-v_\ell}.   \nonumber\\
\]
The right-hand side here is
  increasing in~$v_\ell$, and the left-hand side is increasing in~$v_{\ell-1}$ (recall
	that~$\zeta_{\ell}$ and~$\zeta_{\ell-1}$ are the same for both the original and the modified RFMNP),
so a change in~$\lambda_i$ must lead to~$\sgn(\tilde v_{\ell-1})=\sgn(\tilde v_{\ell})$.
Using~\eqref{eq:eqvvis} again yields
\[
 \zeta_{\ell-2}v_{\ell-2} = \frac{\zeta_\ell v_\ell}{1-v_{\ell-1}},
\]
so~$\sgn(\tilde v_{\ell-2})=\sgn(\tilde v_{\ell-1})=\sgn(\tilde v_{\ell})$.
Continuing in this way yields
\begin{equation}\label{signotherrfm}
   \sgn(\tilde v_{1})=\sgn(\tilde v_2)=\dots =\sgn(\tilde v_{\ell}).
\end{equation}
By \eqref{eq:eqvvis},
\[
      G_2(e_z) = \frac{\zeta_\ell v_\ell}{\zeta_0(1-v_1)}.
\]
Since~$G_i(p)$ is strictly increasing in~$p$, combining this with~\eqref{signotherrfm} implies that
$\sgn(\tilde e_z)=\sgn(\tilde v_{1})=\sgn(\tilde v_2)=\dots =\sgn(\tilde v_{\ell})$.
This proves~\eqref{eq:need2}.
To prove~\eqref{eq:need1}, note that arguing as above using~\eqref{eq:eqeiswithbar} yields
\[
\sgn(\tilde e_n)=\sgn(\tilde e_{n-1})=\dots=\sgn(\tilde e_{i+1}).
\]
Seeking a contradiction, assume that~$\sgn(\tilde e_i)=\sgn(\tilde e_n)$.
By~\eqref{eq:eqeiswithbar},
\[
						e_{i-1}=\frac{\lambda_n e_n}{\lambda_{i-1} (1-e_i) }
\]
so~$\sgn(\tilde e_{i-1})=\sgn(\tilde e_n)$, and continuing in this fashion yields
\be\label{eq:gtr12}
\sgn(\tilde e_n)=\sgn(\tilde e_{n-1})=\dots=\sgn(\tilde e_{1}).
\ee
By~\eqref{eq:eqeis},
\[
G_1(e_z) = \frac{\lambda_n e_n} {\lambda_0(1-e_1)},
\]
and combining this with~\eqref{eq:gtr12} implies that~$\sgn(\tilde e_1)=...=\sgn(\tilde e_n)=\sgn(\tilde e_z)$.
Since also~$\sgn(\tilde v_1)=...=\sgn(\tilde v_\ell)=\sgn(\tilde e_z)$, it  follows that \emph{all} the
differences have the same sign, and this contradicts~\eqref{eq:fiz:totalh}.
We conclude  that if~$\tilde e_i\not = 0$ then~$\sgn(\tilde e_i) \not= \sgn(\tilde e_{i+1}) =\sgn(\tilde e_{i+2})
=\dots  = \sgn(\tilde e_n)$.
To complete the proof of~\eqref{eq:need1}, we need to show that~$\tilde e_i<0$.
  Seeking a contradiction, assume that~$\tilde e_i \geq 0$, so~$\tilde e_{i+1}\leq 0,
	\dots, \tilde{e_n} \leq 0$.
Thus,
\[
							\bar e_i\geq e_i,\quad \bar e_{i+1} \leq  e_{i+1},\dots, \bar e_n\leq e_n.
\]
By~\eqref{eq:eqeis},
$
\lambda_i = \frac{\lambda_n e_n}{e_i(1-e_{i+1})}
$,
so
\begin{align*}
						\frac{		\bar	\lambda_i-\lambda_i }{\lambda_n}
						     &=  \frac{ \bar e_n}{\bar e_i(1-\bar e_{i+1})}  - \frac{ e_n}{e_i(1-e_{i+1})}  \\
								&\leq 0.
\end{align*}
This contradiction completes the proof for the case where~$i\in\{1,\dots,n-1\}$. The proof for the case~$i=0$ and~$i=n$ is similar.~$\square$

\textcolor{black}{
\noindent {\sl Proof of Proposition~\ref{prop:different}.}
In~\cite{Mierc1991},
Mierczynski  considered an irreducible  cooperative dynamical system,~$\dot x=f(x)$, that admits
   a non trivial first integral~$H(x)$
with  a positive gradient. Let~$S_x:=\{v\in\R^n: \nabla H^T(x)v=0 \}$,
and consider the extended  system
\begin{align*}
\dot x&=f(x),\\
\dot {\delta x}&= J(x)\delta x,
\end{align*}
where~$J:=\frac{\partial }{\partial x} f$ is the Jacobian, with initial condition $x(0)=x_0$, $\delta x(0)=\delta x_0 \in S_x \setminus \{0\}$ .
Mierczynski shows that there exists a norm, that depends on~$x$, such that
\[
                    |\delta x(t)|_{x(t )} < |\delta x_0|_{x_0},\quad \text{for all }t>0.
\]
(For a   general treatment on  using Lyapunov-Finsler functions in contraction theory, see~\cite{contra_sep}.)
At the unique equilibrium point~$e$ this yields
\[
|\exp( J(e)t) \delta x_0 |_e < |\delta x_0|_e   ,\quad \text{for all }t>0.
\]
This implies that the matrix  obtained by restricting~$J(e)$   to the integral manifold
 is Hurwitz  and thus
  nonsingular.
  Invoking  the implicit function theorem  implies that the mapping from~$\lambda$ to~$e$
  can be identified with a~$C^1$ function.~$\square$
}

\bibliographystyle{IEEEtranS}

 \end{document}

%% file: fig_rfm_l0.pstex_t
\begin{picture}(0,0)%
\includegraphics{fig_rfm_l0.pstex}%
\end{picture}%
\setlength{\unitlength}{3947sp}%
\begingroup\makeatletter\ifx\SetFigFont\undefined%
\gdef\SetFigFont#1#2#3#4#5{%
  \reset@font\fontsize{#1}{#2pt}%
  \fontfamily{#3}\fontseries{#4}\fontshape{#5}%
  \selectfont}%
\fi\endgroup%
\begin{picture}(7356,1860)(1042,-676)
\put(7276,-136){\makebox(0,0)[lb]{\smash{{\SetFigFont{14}{16.8}{\rmdefault}{\mddefault}{\updefault}$R(t)=\lambda_n x_n(t)$}}}}
\put(7276,239){\makebox(0,0)[lb]{\smash{{\SetFigFont{14}{16.8}{\rmdefault}{\mddefault}{\updefault}Production}}}}
\put(7276,464){\makebox(0,0)[lb]{\smash{{\SetFigFont{14}{16.8}{\rmdefault}{\mddefault}{\updefault}Protein}}}}
\put(1276,989){\makebox(0,0)[lb]{\smash{{\SetFigFont{14}{16.8}{\rmdefault}{\mddefault}{\updefault}$\lambda_0$}}}}
\put(2176,989){\makebox(0,0)[lb]{\smash{{\SetFigFont{14}{16.8}{\rmdefault}{\mddefault}{\updefault}$\lambda_1$}}}}
\put(3076,989){\makebox(0,0)[lb]{\smash{{\SetFigFont{14}{16.8}{\rmdefault}{\mddefault}{\updefault}$\lambda_2$}}}}
\put(3976,989){\makebox(0,0)[lb]{\smash{{\SetFigFont{14}{16.8}{\rmdefault}{\mddefault}{\updefault}$\lambda_3$}}}}
\put(5776,989){\makebox(0,0)[lb]{\smash{{\SetFigFont{14}{16.8}{\rmdefault}{\mddefault}{\updefault}$\lambda_{n-1}$}}}}
\put(6676,989){\makebox(0,0)[lb]{\smash{{\SetFigFont{14}{16.8}{\rmdefault}{\mddefault}{\updefault}$\lambda_n$}}}}
\put(1651,-586){\makebox(0,0)[lb]{\smash{{\SetFigFont{14}{16.8}{\rmdefault}{\mddefault}{\updefault}$x_1(t)$}}}}
\put(2551,-586){\makebox(0,0)[lb]{\smash{{\SetFigFont{14}{16.8}{\rmdefault}{\mddefault}{\updefault}$x_2(t)$}}}}
\put(3451,-586){\makebox(0,0)[lb]{\smash{{\SetFigFont{14}{16.8}{\rmdefault}{\mddefault}{\updefault}$x_3(t)$}}}}
\put(6151,-586){\makebox(0,0)[lb]{\smash{{\SetFigFont{14}{16.8}{\rmdefault}{\mddefault}{\updefault}$x_n(t)$}}}}
\put(5101,-661){\makebox(0,0)[lb]{\smash{{\SetFigFont{14}{16.8}{\rmdefault}{\mddefault}{\updefault}Codon}}}}
\put(4426,-361){\makebox(0,0)[lb]{\smash{{\SetFigFont{14}{16.8}{\rmdefault}{\mddefault}{\updefault}Site \#3}}}}
\end{picture}%

%% file: fig_rfm_network_with_pool.pstex_t
\begin{picture}(0,0)%
\includegraphics{fig_rfm_network_with_pool.pstex}%
\end{picture}%
\setlength{\unitlength}{2486sp}%
\begingroup\makeatletter\ifx\SetFigFont\undefined%
\gdef\SetFigFont#1#2#3#4#5{%
  \reset@font\fontsize{#1}{#2pt}%
  \fontfamily{#3}\fontseries{#4}\fontshape{#5}%
  \selectfont}%
\fi\endgroup%
\begin{picture}(11319,5829)(-1991,-5788)
\put(-809,-2986){\makebox(0,0)[lb]{\smash{{\SetFigFont{10}{12.0}{\rmdefault}{\mddefault}{\updefault}$POOL$}}}}
\put(6121,-5416){\makebox(0,0)[lb]{\smash{{\SetFigFont{10}{12.0}{\rmdefault}{\mddefault}{\updefault}$y^m$}}}}
\put(6121,-2491){\makebox(0,0)[lb]{\smash{{\SetFigFont{10}{12.0}{\rmdefault}{\mddefault}{\updefault}$y^2$}}}}
\put(6121,-961){\makebox(0,0)[lb]{\smash{{\SetFigFont{10}{12.0}{\rmdefault}{\mddefault}{\updefault}$y^1$}}}}
\put(3061,-2491){\makebox(0,0)[lb]{\smash{{\SetFigFont{10}{12.0}{\rmdefault}{\mddefault}{\updefault}$u^2$}}}}
\put(3106,-5461){\makebox(0,0)[lb]{\smash{{\SetFigFont{10}{12.0}{\rmdefault}{\mddefault}{\updefault}$u^m$}}}}
\put(3061,-961){\makebox(0,0)[lb]{\smash{{\SetFigFont{10}{12.0}{\rmdefault}{\mddefault}{\updefault}$u^1$}}}}
\put(1351,-4426){\makebox(0,0)[lb]{\smash{{\SetFigFont{10}{12.0}{\rmdefault}{\mddefault}{\updefault}$G_m(z)$}}}}
\put(1711,-2446){\makebox(0,0)[lb]{\smash{{\SetFigFont{10}{12.0}{\rmdefault}{\mddefault}{\updefault}$G_2(z)$}}}}
\put(1351,-1771){\makebox(0,0)[lb]{\smash{{\SetFigFont{10}{12.0}{\rmdefault}{\mddefault}{\updefault}$G_1(z)$}}}}
\end{picture}%